# Spatially distributed computation in cortical circuits

Sergei Gepshtein,[1,2,*] Ambarish Pawar,[1] Sunwoo Kwon,[3] Sergey Savel'ev,[4] Thomas D. Albright[1]

1.  Systems Neurobiology Laboratories, Salk Institute for Biological Studies, La Jolla, CA, USA
2.  Center for Spatial Perception and Concrete Experience, University of Southern California, Los Angeles, CA, USA
3.  School of Optometry, University of California, Berkeley, CA, USA
4.  Department of Physics, Loughborough University, Loughborough, UK

* sergei@salk.edu

## Abstract

The traditional view of neural computation in the cerebral cortex holds that sensory neurons are specialized, i.e., selective for certain dimensions of sensory stimuli.  This view was challenged by evidence of contextual interactions between stimulus dimensions in which a neuron's response to one dimension strongly depends on other dimensions. Here we use methods of mathematical modeling, psychophysics, and electrophysiology to address shortcomings of the traditional view.  Using a model of a generic cortical circuit, we begin with the simple demonstration that cortical responses are always distributed among neurons, forming characteristic waveforms, which we call neural waves.  When stimulated by patterned stimuli, circuit responses arise by interference of neural waves. Resulting patterns of interference depend on interaction between stimulus dimensions.  Comparison of these modeled responses with responses of biological vision makes it clear that the framework of neural wave interference provides a useful alternative to the standard concept of neural computation.

## Teaser

Investigating interference of neural waves helps to overcome limitations of the traditional view of cortical computation.

# Introduction

The traditional view of neural computation in the cerebral cortex holds that sensory neurons are each characterized by selectivity for certain dimensions of sensory stimuli, such as their orientation, direction of movement, and spatiotemporal frequencies.  This view has been augmented by evidence of nonlinear "contextual" interactions between stimulus dimensions (*1–8*).  A striking example of the latter is selectivity of neurons in cortical area MT of alert monkeys (*9*) where neuronal preferences for stimulus spatial frequency were found to depend on stimulus luminance contrast, more so when luminance contrast was a slowly varying function of time. Here we propose a mechanism that can account for this form of contextual influence on neuronal selectivity.

To study this mechanism, we developed a spatially distributed model of cortical computation in which an excitatory-inhibitory circuit (*10, 11*) is used as a motif repeated across neural tissue and connected to form a chain.  We show that responses of this model are always distributed among neurons, forming characteristic waveforms which we term neural waves.  Waves elicited by different parts of the stimulus spread across the chain and interfere with one another. Patterns of interference created by this process have a number of characteristics that determine the circuit's preference for stimulation.  Notably, the circuit's preference for stimulus spatial frequency predicted by this method is expected to depend on both luminance contrast and temporal properties of the stimulus, resulting in patterns of interaction between stimulus dimensions that are strikingly similar to those found in MT neurons, as described above (*9*).

By investigating other properties of interference patterns, we make additional novel predictions about selectivity of cortical circuits and about contextual interaction between stimuli.  We confirm these predictions using physiological and behavioral methods.  We show in particular that neuronal preferences for stimulus temporal frequency depends on stimulus spatial frequency, which is a hitherto unknown phenomenon predicted by the model.  We show also that the modulated activity that arises in the visual system outside of the region of direct stimulation does not depend upon the spatial frequency of the visual stimulus, consistent with the model prediction about interference of neural waves.

These findings suggest that the framework of neural wave interference offers a useful predictive account of neural computation in visual cortical circuits, which we offer for consideration as an alternative to the account of computation by neurons characterized by sustained specialization.

More generally, against the backdrop of increasing interest in wave phenomena in neural systems (*12–14*), where uncertainty lingers about the computational role of such waves (*15–17*), we present a detailed illustration of how neural waves could underlie a basic visual computation.

# Results

Our results consist of two threads, theoretical and empirical.  In the theoretical thread, we examine spatial interactions between neurons using a distributed model of a generic cortical circuit.  In the empirical thread, we test several predictions of this model using sensory psychophysics in human subjects and single-neuron physiology in alert macaque monkeys.

## Model of distributed circuit

### Approach

We developed a spatially distributed model of cortical computation based on the excitatory-inhibitory circuit introduced by Wilson & Cowan (*10, 11*).  Their original circuit is illustrated in Figure 1A.  It consists of excitatory ($E$) and inhibitory ($I$) parts connected reciprocally and recurrently. Extensive studies of this circuit suggested that it can serve as a generic precursor for dynamic models of biological neural networks (*8, 18–24*).

In our model, this circuit is used as a motif. The motifs are repeated across neural tissue and connected to form a chain: a spatially distributed form of the basic excitatory-inhibitory circuit.  In this minimal model we only consider a one-dimensional arrangement of motifs since it provides a simple preparation for investigating spatially distributed phenomena. We consider complete connectivity between neighboring motifs (Figure 1B), in which the excitatory and inhibitory cells of each motif are each connected to both excitatory and inhibitory cells of neighboring motifs.

An emergent property of this spatially distributed circuit is an intrinsic preference for the distributed stimuli that serve as input to the system.  We begin by illustrating the mechanism that underlies this intrinsic preference, in three steps.  First, we use a localized "point" stimulus that activates a single motif and elicits a spatially distributed response: a "standing" neural wave.  Second, we consider two spatially-distinct point stimuli that activate two motifs, which are separated by distances that allow for interference between the resulting neural waves.  Third, we investigate circuit responses to spatially distributed stimuli and compare model predictions in linear and nonlinear regimes with results of psychophysical and physiological studies of biological vision.  We find that the model of a fully connected circuit can successfully explain phenomena of lateral interaction between stimuli and stimulus selectivity of cortical circuits.

### Neural wave interference

We begin by considering a very small ("point") stimulus that initially activates a single motif in the chain.  The point activation propagates through the chain by means of inter-motif coupling.  The steady state of this process is a distributed spatial pattern of activation.  We found the shape of this pattern by integrating the equations in which the circuit is represented as a continuous system (Methods, Equation 2).  The solution can be usefully described as a neural wave with two components: excitatory and inhibitory. Both components are distributed through the network. Both components have the form of a damped wave with the same spatial frequency and the same rate of decay (damping).  An example of the excitatory component is shown in Figure 1C; it is a standing spatially-damped wave (Methods, Equation 3).  The components interact, making them parts of a single neural wave.

A larger stimulus activates multiple parts of the network.  The neural waves originating in different locations in the network form a complex distributed pattern.  Assuming linearity (an assumption we relax in the last section of Results), the resulting pattern can be predicted by the principle of superposition, where the distributed activation is the sum of multiple local waves, forming a pattern of spatial interference.  Activation is expected to be higher or lower depending on whether the interference of waves is constructive or destructive (illustrated below; Figure 1D).  We use the intuition of linear interference of neural waves to illustrate how the distributed circuit becomes intrinsically selective for stimulus spatial frequency (SF).

Suppose the chain is activated by two identical point stimuli, each generating a neural wave originating at a unique motif.  By virtue of interference of these waves, circuit activity between the point stimuli is facilitated or suppressed at different locations.  Figure 1D is a map of the excitatory component of the neural wave distributed over multiple locations in the chain (represented on the abscissa) for different distances $D$ between the point stimuli (represented on the ordinate).  Notice that the response pattern is periodic, but the period does not depend on the inter-stimulus distance.  This period is determined by the weights of excitatory and inhibitory connections between cells; it is an intrinsic property of the circuit (henceforth "intrinsic period" or its inverse: "intrinsic frequency").

The variety of resulting interference patterns can be seen by examining horizontal slices through the map, as shown at right in Figure 1D.  When the two point stimuli overlap at the zeroth location ($D$=0, bottom inset) the positive activation is highest at the location of these stimuli, forming a waveform of positive and negative activations at other locations. Increasing distance $D$ (moving upwards on the ordinate) leads to alternating states of network activity.  When $D$ is proportional to the intrinsic period of the chain, overall activation is heightened because the waves of activation from the two stimulated locations "arrive" in phase with one another, as in the third and fifth insets from bottom where the stimuli are separated by respectively 1 and 2 intrinsic periods.  When distance $D$ is proportional to half the intrinsic period, by contrast, the overall activation is reduced because the waves "arrive" out of phase and cancel one another, as in the second and fourth insets from bottom, where the stimuli are separated by respectively 0.5 or 1.5 intrinsic periods.

### Direct and lateral activation in the model circuit

Next, we investigated the response of the model to distributed spatial stimuli.  In this case, the neural waves elicited on multiple cells in the chain interfere and form a distributed pattern, as they do for the simpler two-point stimulation described above.  Here we considered a common situation in which the stimulus is spatially continuous over a segment of the chain, but it is bounded: stopped abruptly by a well-defined edge.  In the top row of Figure 2 we plot three examples of such a stimulus: a luminance grating at three spatial frequencies (SF). The analytical solution for the model response to each stimulus can be found by convolving the stimulus with the model response to the point stimulus.  The excitatory component of the solution (Methods, Equation 3) is plotted in the bottom row of Figure 2.  This analysis revealed that response properties are qualitatively different in the region of the model circuit where the cells are activated directly (henceforth "Zone 1") and outside of that region ("Zone 2").

**Zone 1.** Well within Zone 1 (shown in Figure 2 in black), the SF of the response waveform is equal to the stimulus SF ($SF_{stim}$).  Near the boundary of Zone 1, but still inside of it, we find a mixture of two frequencies: the $SF_{stim}$ and the intrinsic SF of the system (SF*).  Away from the boundary, response amplitude depends on how close the $SF_{stim}$ is to SF*.  The amplitude is maximal when the $SF_{stim}$ approaches the SF* (spatial resonance).

**Zone 2.** Just outside of the region of direct stimulation, we find a "lateral activation" of the circuit (shown in red).  Here, the response SF changes abruptly to SF*, but the amplitude of response decreases gradually as a function of distance away from the boundary of direct stimulation.  The SF of the response in Zone 2 is independent of the stimulus SF.

In the following sections we evaluate the suitability of the model by empirically testing some of its key predictions.  In the next section, we use psychophysical methods to test the model prediction that the SF of lateral activation in Zone 2 does not depend on the stimulus SF.  In subsequent sections, we examine predictions concerned with circuit's stimulus selectivity in Zone 1.

## Lateral activation in human vision

Numerous psychophysical studies have found that visual systems change their sensitivity to stimulation outside of regions of direct stimulation (*25–28*).  Such results are often attributed to long-range "horizontal" connections between cortical neurons (*29–31*).  With rare exception (*32*), lateral activation is probed using spatially distributed stimuli, such as "Gabor patches."  Responses to distributed probes arise by integration of multiple neural waves, reflecting the structure of the probe in addition to the neural waves generated by other stimuli (Methods, Equation 4).  To avoid this problem, we obtained psychophysical measurements of lateral activation using a small probe.

The stimulus in our psychophysical study consisted of three parts: two inducing luminance gratings ("inducers") and a faint vertical line ("probe") positioned between the inducers (Figure 3A-B).  The inducers were square patches of luminance gratings of fixed luminance contrast, whose abrupt edges ensured that the region of direct stimulation had a well-defined boundary (as in Figure 2) and no direct subthreshold activation of the visual system was introduced outside of the visible part of the stimulus.  We measured contrast sensitivity for the probe, which was placed between the inducers following the procedure described in Methods.  The pattern of contrast sensitivity serves as an index of lateral activation.

Results of this experiment for two human subjects are displayed in Figure 3C-D.  The change in subjects' contrast sensitivity to the probe is plotted as a function of distance from the nearest inducer edge for three values of inducer SF.  The columns of plots contain data for different values of the inducer SF, and the rows contains data for different subjects.

The results displayed in Figure 3 reveal a clear modulation of contrast sensitivity outside of the region of direct stimulation.  These results are evidence that lateral activation of the human visual system yields a pronounced spatial waveform and that the period of this lateral spatial modulation does not systematically depend on the stimulus SF (see Supplemental Figure 1), consistent with the prediction of our model.  The results support the view that lateral activation of the visual system is constituted by neural waves.

## Selectivity of cortical circuits

In addition to eliciting lateral activity, neural waves in the model interact inside the region of direct stimulation, forming interference patterns that define the conditions of circuit resonance, at which the amplitude of response is maximal.  The resonance spatial frequency (SFR) of the circuit has a value close to its intrinsic frequency (SF*) when wave decay over distance (spatial damping) in the system is low (Methods, Equation 16).  We studied conditions of circuit resonance two ways. First, we took into account the fact that our stimuli are modulated in time and thus are characterized by both spatial and temporal frequencies (SF and TF).  Using a linear approximation of the model, we investigated the conditions of spatial and temporal resonance that determine, respectively, spatial and temporal stimulus preferences of the circuit.  Second, we investigated how conditions of resonance depend on stimulus contrast, which required that we study circuit behavior beyond linear approximation.

### Spatiotemporal selectivity in the model circuit

**Spatial selectivity.** Results of our analysis of SFR in the model (Methods, Equations 6-9) are summarized in Figures 4 and 5A.  We found that increasing stimulus TF leads to changes in spatial resonance (SFR).  The form of this relationship depends on the weights of connections in the circuit, forming two qualitatively different patterns.  In one pattern (which we call Regime 1) SFR increases with TF; in the other pattern (Regime 2) SFR decreases with TF.

Regime 1 is illustrated in Figure 4A, where model responses are plotted for different values of stimulus TF, represented by separate curves.  The peak of each response function is found at the SFR for the indicated stimulus TF.  Response peaks shift <u>rightward</u>, toward higher SF values, as TF grows, represented in Figure 5A by the black curve at left. Regime 2 is illustrated in Figure 4B.  Here, response peaks shift <u>leftward</u> (toward lower SF values) as TF grows, represented in Figure 5A by the red curve at right.

The rise of SFR in Regime 1 is bounded by a vertical asymptote (a separatrix) represented in Figure 5A by a dashed vertical line.  The decline of SFR in Regime 2 is bounded by the same vertical asymptote. In summary, as TF increases, SFR approaches the asymptote from the left side in Regime 1 and from the right side in Regime 2.  As weights change, system behavior changes continuously from one regime to another, passing through conditions in which SFR does not depend on TF.  This remarkable result also suggests a different explanation, in terms of interaction of excitatory and inhibitory components of the neural wave rather than weights between cells.  From the perspective of interaction of wave components, different regimes of circuit behavior (and change of circuit behavior from one regime to another) can arise because coefficients of interaction between neural wave components depend on stimulus TF (Methods, Equation 6).

**Temporal selectivity.**  Next we show how temporal resonance of the model (its resonance TF, abbreviated TFR) depends on stimulus SF.  In this case we vary stimulus TF for fixed values of stimulus SF to find the TF of maximal response (TFR). We perform this analysis of TFR for the same parameters of the model circuit at which we have obtained the solutions for SFR in the preceding section (Methods, Equation 9).  The predicted pattern of TFR (Figure 5B) is radically different from the pattern of SFR in terms of the shape of interaction of spatial and temporal stimulus frequencies (Methods, Equation 10).  The prediction is that TFR should remain a nonmonotonic function of stimulus SF.

### *Spatiotemporal selectivity in biological cortical circuits*

We tested predictions of our network model with regard to spatial and temporal frequency selectivity in biological cortical circuits.  The predictions summarized in Figure 5A-B were tested by applying new analyses to data that were obtained in a previously reported (*9*) physiological study of stimulus preferences in the middle temporal (MT) cortical area of two awake macaque monkeys.  A preview of the physiological results appears in Figure 5C-F.  The juxtaposition of theoretical and physiological results in Figure 5 reveals a striking similarity of changes in system preference predicted by the model and measured in the monkey.

The stimuli used in the physiological study were drifting luminance gratings in which we varied SF, TF, and luminance contrast (henceforth "contrast"). Following the procedure described in Methods, we measured the firing rates of neurons for several values of stimulus SF at fixed stimulus TF and contrast, allowing us to estimate the neuron's preference for stimulus SF. We then repeated this procedure for different fixed values of TF. By this means we characterized each neuron's spatial and temporal stimulus preferences for several values of stimulus contrast. We obtained such data from 74 neurons in Monkey 1 and 66 neurons in Monkey 2. In Figures 6-7, we present results of an analysis of these data aggregated across all recorded neurons, separately for each monkey.

*Temporal selectivity*. Effects of stimulus SF on temporal selectivity of cortical neurons are illustrated in Figure 6. Each plot in the top row of panel A represents average neuronal firing rate (ordinate) for Monkey 1 as a function of stimulus TF (abscissa), for a single value of stimulus SF (indicated at top left in every plot) and a single stimulus contrast of 7%. By fitting the data across stimulus TF, we estimated the preferred TF for each stimulus SF, represented in each plot by a vertical line. A summary of these estimates for the stimulus contrast of 7% appears in panel B, revealing how the preferred TF of the population of cortical neurons varies with stimulus SF. Results of the same analyses for two other stimulus contrasts in Monkey 1 are displayed in panels A and C, and results for the same three stimulus contrasts in Monkey 2 are displayed in panels D-F. In both monkeys, the preferred TF follows a nonmonotonic function of stimulus SF. The shape of this function does not change with stimulus contrasts (confirmed for most conditions as detailed in the caption of Figure 6), following the pattern predicted by our model (Figure 5B).

*Spatial selectivity*. Effects of stimulus TF on spatial selectivity of cortical neurons are summarized in Figure 7. Each plot in the left column of panels represents average neuronal firing rate (ordinate) for Monkey 1 as a function of stimulus SF (abscissa), for a single value of stimulus TF (indicated at top left in every plot) and a single stimulus contrast of 2%. By fitting the data across stimulus SF, we estimated the preferred SF for each stimulus TF, represented in each plot by a vertical line. A summary of these estimates for the stimulus contrast of 2% appears in panel A, together with estimates of preferred SF for two other stimulus contrasts. Results of the same analyses of data for Monkey 2 are displayed in panel B.

The plots in Figures 7A-B indicate that increasing stimulus contrast led to a significant increase in the peak SF in both monkeys: for TFs below 8 Hz in Monkey 1 and for all TFs in Monkey 2. Additionally, in both monkeys peak SF increased with TF at low stimulus contrast and decreased with TF at high stimulus contrast.  This pattern of change of preferred SF of cortical neurons is similar to model predictions (Figure 5A) in which the shape of the SFR function changes radically: from an increasing function of TF in one regime to a decreasing function of TF in the other regime, while these functions converge with increasing TF.

In our modeling framework, these changes of cortical selectivity represent different instances of the same general pattern of dependency of SFR on stimulus TF.  Beyond model predictions, the physiological results reveal that the separation of two regimes of interaction between peak SF and stimulus TF can be caused by stimulus contrast.

### *Stimulus contrast and cortical selectivity*

Our physiological finding that the regime of system preference for SF depends on stimulus contrast suggests that our linear analysis of the circuit was too narrow.  We therefore studied how system nonlinearity affects conditions of circuit resonance (Equations 11-17).  This analysis made it clear that resonance frequency of the circuit depends on the interaction of excitatory and inhibitory components of neural waves.  At low contrast, in the linear regime, the interaction of wave components depends on weights of cell connections and on stimulus spatial and temporal frequencies.  At higher contrasts, however, the interaction becomes nonlinear, and it additionally depends on the amplitudes of the neural wave themselves (Equation 15).  It is because the interaction between excitatory and inhibitory components of the neural wave depends on stimulus contrast that resonance SF can increase or decrease with contrast.

To illustrate this neural wave interaction, we plot in Figure 8A the amplitude of the excitatory wave component as a function of both stimulus SF and contrast.  Vertical slices through the surface, orthogonal to the contrast axis, correspond to individual SF response functions for different levels of stimulus contrast.  Five of these slices are shown in Figure 8B.  In each case, responses follow a smooth unimodal function of SF.  The maximum of each function identifies the resonance SF (SFR) for the corresponding contrast.  At low contrasts, where circuit behavior is approximately linear, SFR does not change with contrast (red and green curves).

Increasing contrast leads to a uniformly higher activation of the circuit and, notably, to a change of SFR: its gradual increase (Figure 8A-B) or decrease (Figure 8C).

Neuronal preferences for stimulus SF in area MT were shown to change with contrast (*9*) (also see Figure 7 above). In an additional analysis, we derived neuronal population response functions using the aggregate estimates of firing rates, as in Figures 6-7.  These functions are plotted separately for four values of stimulus contrast in Figure 8D-E.  In both monkeys, the mean preferred SF increased with contrast, as shown at left of Figure 8F: peak SF increased 3.11 fold in Monkey 1 and 2.14 fold in Monkey 2.  This increase of peak SF in populations of neurons was accompanied by changes in individual-cell preferences summarized by six donut charts at right of Figure 8F.  For example, the chart at top right indicates that increasing stimulus contrasts from 14% to 52% in Monkey 1 led to significantly increasing peak SF in 69% of cases, significantly decreasing peak SF in 23% of cases, and caused no significantly change in 8% of cases.  The fraction of neurons in which peak SF increased with contrast itself increased with contrast, as evident by comparing the top and bottom panels at right of Figure 8F.  In summary, we found that peak SF changed with contrast in a large majority of cortical neurons, consistent with the prediction of our model that resonance SF can increase or decrease with stimulus contrast.  In the model, SFR increases or decreases with stimulus contrast because the interaction between excitatory and inhibitory components of the neural waves depends on stimulus contrast.

## Discussion

We studied spatially distributed cortical computations by means of mathematical modeling, human psychophysics, and physiological recordings from isolated cortical neurons in non-human primates.  We proposed a distributed model of cortical circuits that can account for some puzzling properties of the primate visual system and also predict properties not heretofore examined.  Here we discuss the possibility that interference of neural waves provides a general framework for understanding cortical computation.

We observed at the outset that point stimulation of the model circuit elicits a distributed periodic response: a neural wave with two components, excitatory and inhibitory, that interact with one another.  The spatiotemporal frequency of the neural wave is an intrinsic property of the circuit. When stimulation is distributed (i.e., applied to multiple points in the circuit), as is generally the case in biological vision, the neural waves evoked at different locations interfere with one another.  This pattern of interference determines the stimuli to which the system resonates, thus defining the stimulus selectivity of the model.  Neural waves also propagate outside of the region of direct stimulation, causing lateral activation of the system.  These two behaviors, concerned with system selectivity and its lateral activation, have been previously studied in the framework of separate mechanisms of classical and nonclassical receptive fields (*33–40*).

To assess the validity of this model we empirically tested its specific predictions regarding stimulus selectivity and lateral activation in the primate visual system using physiological and psychophysical methods.

Using physiological methods, we found that the changes of stimulus preference in biological cortical circuits depend on input stimulus parameters in a manner similar to such dependences in the model.  In addition to our previous discovery that SF selectivity depends upon stimulus contrast and temporal frequency (*9*), we report here that selectivity for stimulus temporal frequency depends upon stimulus spatial frequency, which is a hitherto unknown phenomenon predicted by the model.  Using psychophysical methods, we found that contrast sensitivity outside of the region of direct stimulation forms a modulated pattern whose spatial frequency is independent of stimulus spatial frequency.  This finding supports the notion that lateral modulation of contrast sensitivity reflects an intrinsic neural wave.  Taken together, our results suggest that the generic model of the distributed excitatory-inhibitory circuit offers a useful account of biological vision, and that the language of interaction of neural waves affords explanation of phenomena of cortical selectivity that have resisted explanation in terms of specialized cortical neurons.

The model of wave interaction in neural circuits is readily generalizable to two spatial dimensions, suited to study cortical selectivity for stimulus orientation (*41*, *42*), shape (*26*, *28*, *43*, *44*) and stereoscopic depth (*45*, *46*).  It is equally suited to study temporal interactions between successive stimuli (*27*, *47*, *48*) and the interaction between spatial and temporal dynamics of cortical activity (*49*, *50*), in cortical area MT and other cortical areas.  Two examples of generalization of our model, spatial and temporal, are illustrated in Figures 9-10.

In Figure 9 we illustrate interference of neural waves in two spatial dimensions.  We obtained these results by means of numerical simulation of activity in a two-dimensional lattice of nodes (Supplemental Modeling Methods), where each node was the same excitatory-inhibitory circuit as in our one-dimensional model (Figure 1A).  The nodes were connected along the sides and diagonals of the lattice, following the same pattern of connectivity of excitatory and inhibitory cells as in Figure 1B.  Neural waves originating in different locations in the network form a complex pattern of spatial interference, as in the one-dimensional model, except now this pattern is distributed in two spatial dimensions, even for a point stimulus (Figure 9A). We used a high-contrast stimulus that had the shape of an elliptical ring, represented in Figure 9B by a dashed black line. Interference of neural waves generated by this stimulus produced two salient regions of excitation near the two foci of the elliptical ring, clearly visible in the two-dimensional map of activation in Figure 9B. Contrast threshold in these locations should be lower than in neighboring locations, in agreement with psychophysical findings (Figure 9C) (*26*, *51*).

In Figure 10 we illustrate interference of neural oscillations across time. Similar to how neuronal waves interfere across network locations, temporal oscillations at every location in the network interfere across time. Activity produced at instant $t_0$ will form a damped neural oscillation (e.g., 'a' in Figure 10B) across time $t$-$t_0$. Activity produced at the same location at later instants $t_1$ and $t_2$ will produce additional oscillations ('b' and 'c' in Figure 10B). Such subsequent oscillations superpose with one another, forming patterns of interference whose temporal profiles depend on properties of individual oscillations generated at different instances (Figure 10C). One outcome of temporal interference of neural oscillations in our model is illustrated in Figure 10D. We studied how duration and luminance contrast of a visual stimulus affect its visibility across time. Intuitively, one expects that increasing stimulus duration and luminance contrast should make the stimulus more visible. Our study of interference of neural oscillations has suggested that, paradoxically, increasing stimulus duration can decrease visibility. By numerically integrating responses of a single node of our model (Figure 1A) over time, we investigated *duration threshold*, which is the stimulus duration at which the stimulus should become just visible (Supplemental Modeling Methods). In a system dominated by excitation, increasing contrast led to reduction of duration threshold, helping to make the stimulus more visible (i.e., requiring lower contrast to reach threshold). This result, represented in Figure 10D by the gray curve, agrees with one's intuition. But in a system dominated by inhibition, represented in the figure by the green curve, increasing contrast leads to nonmonotonic changes of duration threshold, which call fall or rise with contrast under different values of contrast. Such properties of temporal interference can help to explain results of studies of duration threshold in human subjects (*47*, *48*, *52*) which revealed a similarly counterintuitive behavior. Duration threshold was found to rise or fall with stimulus contrast depending on stimulus size (Figure 10E), which is known to control the amount of inhibition in cortical circuits (*35*, *36*).

In summary, two illustrations presented in Figures 9-10 indicate that the model of neural wave interference has the potential for application to visual phenomena beyond those reported in the empirical part of this study. Taken together, our findings suggest that cortical computations are mediated by interference of neural waves, at least in part. We have presented a detailed account of how neural waves could underlie basic visual computations, suggesting specific roles for wave phenomena in visual perception. The concept of neural computation by wave interference in cortical circuits can help to overcome limitations of traditional concepts, in which distributed cortical activity is viewed simply in terms of interaction of specialized neurons.

**Materials and Methods.** The section of Materials and Methods is appended to the end of this document because it is typeset and compiled using the software system LaTeX.

**Acknowledgements.** We thank D. Alton, B. Berthet, D. Diep, E. Suljic, and K. Zornado for superb technical assistance. This work was supported by the Sloan-Swartz Center for Theoretical Neurobiology, the Kavli Institute for Brain and Mind (University of California, San Diego), NIH grants R01-EY018613, R01-EY029117, EPSRC grant EP/S032843/1, and the Conrad T. Prebys Chair in Vision Research.

**Author contributions.** Conceptualization: S.G., S.S., T.D.A.; funding acquisition: S.G., S.S., T.D.A.; modeling curation: S.G., S.S.; mathematical analysis: S.S.; curation of psychophysical experiments: S.G., S.K., S.S.; psychophysical data collection: S.K.; psychophysical data analysis: S.G., S.K.; curation of physiological experiments: S.G., A.P.,T.D.A; physiological data collection: A.P.; physiological data analysis: S.G., A.P.; visualization: S.G., A.P., S.K., S.S.; writing original draft, review, editing: S.G., A.P., S.S., T.D.A.

**Competing interests.** The authors declare no competing interests.

# References

1.  R. A. Holub, M. Morton-Gibson, Response of Visual Cortical Neurons of the cat to moving sinusoidal gratings: response-contrast functions and spatiotemporal interactions. *J. Neurophysiol.* **46**, 1244–1259 (1981).

2.  J. Allman, F. Miezin, E. McGuinness, Stimulus specific responses from beyond the classical receptive field: Neurophysiological mechanisms for local-global comparisons in visual neurons. *Annu. Rev. Neurosci.* **8**, 407–430 (1985).

3.  G. C. DeAngelis, R. D. Freeman, I. Ohzawa, Length and width tuning of neurons in the cat's primary visual cortex. *J. Neurophysiol.* **71**, 347–374 (1994).

4.  D. G. Albrecht, Visual cortex neurons in monkey and cat: effect of contrast on the spatial and temporal phase transfer functions. *Vis. Neurosci.* **12**, 1191–1210 (1995).

5.  T. D. Albright, G. R. Stoner, Contextual influences on visual processing. *Annu. Rev. Neurosci.* **25**, 339–379 (2002).

6.  M. P. Sceniak, M. J. Hawken, R. Shapley, Contrast-Dependent Changes in Spatial Frequency Tuning of Macaque V1 Neurons: Effects of a Changing Receptive Field Size. *J. Neurophysiol.* **88**, 1363–1373 (2002).

7.  N. J. Priebe, S. G. Lisberger, J. A. Movshon, Tuning for spatiotemporal frequency and speed in directionally selective neurons of macaque striate cortex. *J. Neurosci.* **26**, 2941–2950 (2006).

8.  D. B. Rubin, S. D. Van Hooser, K. D. Miller, The stabilized supralinear network: A unifying circuit motif underlying multi-input integration in sensory cortex. *Neuron.* **85**, 402–417 (2015).

9.  A. S. Pawar, S. Gepshtein, S. Savel'ev, T. D. Albright, Mechanisms of Spatiotemporal Selectivity in Cortical Area MT. *Neuron.* **101**, 514-527.e2 (2019).

10. H. R. Wilson, J. D. Cowan, Excitatory and inhibitory interactions in localized populations of model neurons. *Biophys. J.* **12**, 1–24 (1972).

11. H. R. Wilson, J. D. Cowan, A mathematical theory of the functional dynamics of cortical and thalamic nervous tissue. *Kybernetik.* **13**, 55–80 (1973).

12. M. G. Preti, D. Van De Ville, Decoupling of brain function from structure reveals regional behavioral specialization in humans. *Nat. Commun.* **10**, 4747 (2019).

13. J. A. Roberts, L. L. Gollo, R. G. Abeysuriya, G. Roberts, P. B. Mitchell, M. W. Woolrich, M. Breakspear, Metastable brain waves. *Nat. Commun.* **10**, 1056 (2019).

14. Z. W. Davis, L. Muller, J. Martinez-Trujillo, T. Sejnowski, J. H. Reynolds, Spontaneous travelling cortical waves gate perception in behaving primates. *Nature*, 1–5 (2020).

15. I. Nauhaus, L. Busse, M. Carandini, D. L. Ringach, Stimulus contrast modulates functional connectivity in visual cortex. *Nat. Neurosci.* **12**, 70–76 (2009).

16. L. Muller, F. Chavane, J. Reynolds, T. J. Sejnowski, Cortical travelling waves: mechanisms and computational principles. *Nat. Rev. Neurosci.* **19**, 255–268 (2018).

17. S. Hellrigel, N. Jarman, C. van Leeuwen, Adaptive rewiring in weighted networks. *Cogn. Syst. Res.* **55**, 205–218 (2019).

18. M. V. Tsodyks, W. E. Skaggs, T. J. Sejnowski, B. L. McNaughton, Paradoxical Effects of External Modulation of Inhibitory Interneurons. *J. Neurosci.* **17**, 4382–4388 (1997).

19. H. R. Wilson, *Spikes, Decisions, and Actions: The Dynamical Foundations of Neuroscience* (Oxford University Press, USA, 1999).

20. A. F. Teich, N. Qian, Learning and Adaptation in a Recurrent Model of V1 Orientation Selectivity. *J. Neurophysiol.* **89**, 2086–2100 (2003).

21. H. Ozeki, I. M. Finn, E. S. Schaffer, K. D. Miller, D. Ferster, Inhibitory stabilization of the cortical network underlies visual surround suppression. *Neuron.* **62**, 578–592 (2009).

22. Y. Ahmadian, D. B. Rubin, K. D. Miller, Analysis of the stabilized supralinear network. *Neural Comput.* **25**, 1994–2037 (2013).

23. M. P. Jadi, T. J. Sejnowski, Regulating cortical oscillations in an inhibition-stabilized network. *Proc. IEEE.* **102**, 830–842 (2014).

24. K. D. Miller, Canonical computations of cerebral cortex. *Curr. Opin. Neurobiol.* **37**, 75–84 (2016).

25. U. Polat, D. Sagi, Lateral interactions between spatial channels: suppression and facilitation revealed by lateral masking experiments. *Vision Res.* **33**, 993–999 (1993).

26. I. Kovacs, B. Julesz, Perceptual sensitivity maps within globally defined visual shapes. *Nature.* **370**, 644–646 (1994).

27. V. Manahilov, Triphasic temporal impulse responses and Mach bands in time. *Vision Res.* **38**, 447–458 (1998).

28. D. J. Field, A. Hayes, R. F. Hess, Contour integration by the human visual system: evidence for a local association field. *Vision Res.* **33**, 173–193 (1993).

29. C. D. Gilbert, T. N. Wiesel, Columnar specificity of intrinsic horizontal and corticocortical connections in cat visual cortex. *J. Neurosci.* **9**, 2432–2442 (1989).

30. D. D. Stettler, A. Das, J. Bennett, C. D. Gilbert, Lateral Connectivity and Contextual Interactions in Macaque Primary Visual Cortex. *Neuron.* **36**, 739–750 (2002).

31. J. D. Semedo, A. Zandvakili, C. K. Machens, B. M. Yu, A. Kohn, Cortical Areas Interact through a Communication Subspace. *Neuron.* **102**, 249-259.e4 (2019).

32. V. Manahilov, Spatiotemporal visual response to suprathreshold stimuli. *Vision Res.* **35**, 227–237 (1995).

33. J. R. Cavanaugh, W. Bair, J. A. Movshon, Nature and Interaction of Signals from the Receptive Field Center and Surround in Macaque V1 Neurons. *J. Neurophysiol.* **88**, 2530–2546 (2002).

34. W. Bair, Visual receptive field organization. *Curr. Opin. Neurobiol.* **15**, 459–464 (2005).

35. R. T. Born, D. C. Bradley, Structure and Function of Visual Area MT. *Annu. Rev. Neurosci.* **28**, 157–189 (2005).

36. A. Angelucci, M. Bijanzadeh, L. Nurminen, F. Federer, S. Merlin, P. C. Bressloff, Circuits and Mechanisms for Surround Modulation in Visual Cortex. *Annu. Rev. Neurosci.* **40**, 425–451 (2017).

37. D. Fitzpatrick, The Functional Organization of Local Circuits in Visual Cortex: Insights from the Study of Tree Shrew Striate Cortex. *Cereb. Cortex*. **6**, 329–341 (1996).

38. J. B. Levitt, J. S. Lund, in *Cortical Areas* (CRC Press, 2002), pp. 145–166.

39. O. Schwartz, A. Hsu, P. Dayan, Space and time in visual context. *Nat. Rev. Neurosci.* **8**, 522–535 (2007).

40. D. L. Ringach, The geometry of masking in neural populations. *Nat. Commun.* **10**, 4879 (2019).

41. J. J. Pattadkal, G. Mato, C. van Vreeswijk, N. J. Priebe, D. Hansel, Emergent Orientation Selectivity from Random Networks in Mouse Visual Cortex. *Cell Rep.* **24**, 2042-2050.e6 (2018).

42. N. Ju, Y. Li, F. Liu, H. Jiang, S. L. Macknik, S. Martinez-Conde, S. Tang, Spatiotemporal functional organization of excitatory synaptic inputs onto macaque V1 neurons. *Nat. Commun.* **11**, 697 (2020).

43. R. F. Hess, A. Hayes, D. J. Field, Contour integration and cortical processing. *J. Physiol.-Paris*. **97**, 105–119 (2003).

44. D. A. Mély, D. Linsley, T. Serre, Complementary surrounds explain diverse contextual phenomena across visual modalities. *Psychol. Rev.* **125**, 769–784 (2018).

45. H. R. Wilson, Binocular contrast, stereopsis, and rivalry: Toward a dynamical synthesis. *Vision Res.* **140**, 89–95 (2017).

46. G. Riesen, A. M. Norcia, J. L. Gardner, Humans Perceive Binocular Rivalry and Fusion in a Tristable Dynamic State. *J. Neurosci.* **39**, 8527–8537 (2019).

47. D. Tadin, J. S. Lappin, L. A. Gilroy, R. Blake, Perceptual consequences of centre–surround antagonism in visual motion processing. *Nature*. **424**, 312–315 (2003).

48. D. Tadin, Suppressive mechanisms in visual motion processing: From perception to intelligence. *Vision Res.* **115**, 58–70 (2015).

49. D. M. Alexander, T. Ball, A. Schulze-Bonhage, C. van Leeuwen, Large-scale cortical travelling waves predict localized future cortical signals. *PLOS Comput. Biol.* **15**, e1007316 (2019).

50. H. Hogendoorn, Motion Extrapolation in Visual Processing: Lessons from 25 Years of Flash-Lag Debate. *J. Neurosci.* **40**, 5698–5705 (2020).

51. I. Kovacs, B. Julesz, A closed curve is much more than an incomplete one: Effect of closure in figure-ground segmentation. *Proc. Natl. Acad. Sci.* **90**, 7495–7497 (1993).

52. D. Tadin, J. S. Lappin, Optimal size for perceiving motion decreases with contrast. *Vision Res.* **45**, 2059–2064 (2005).

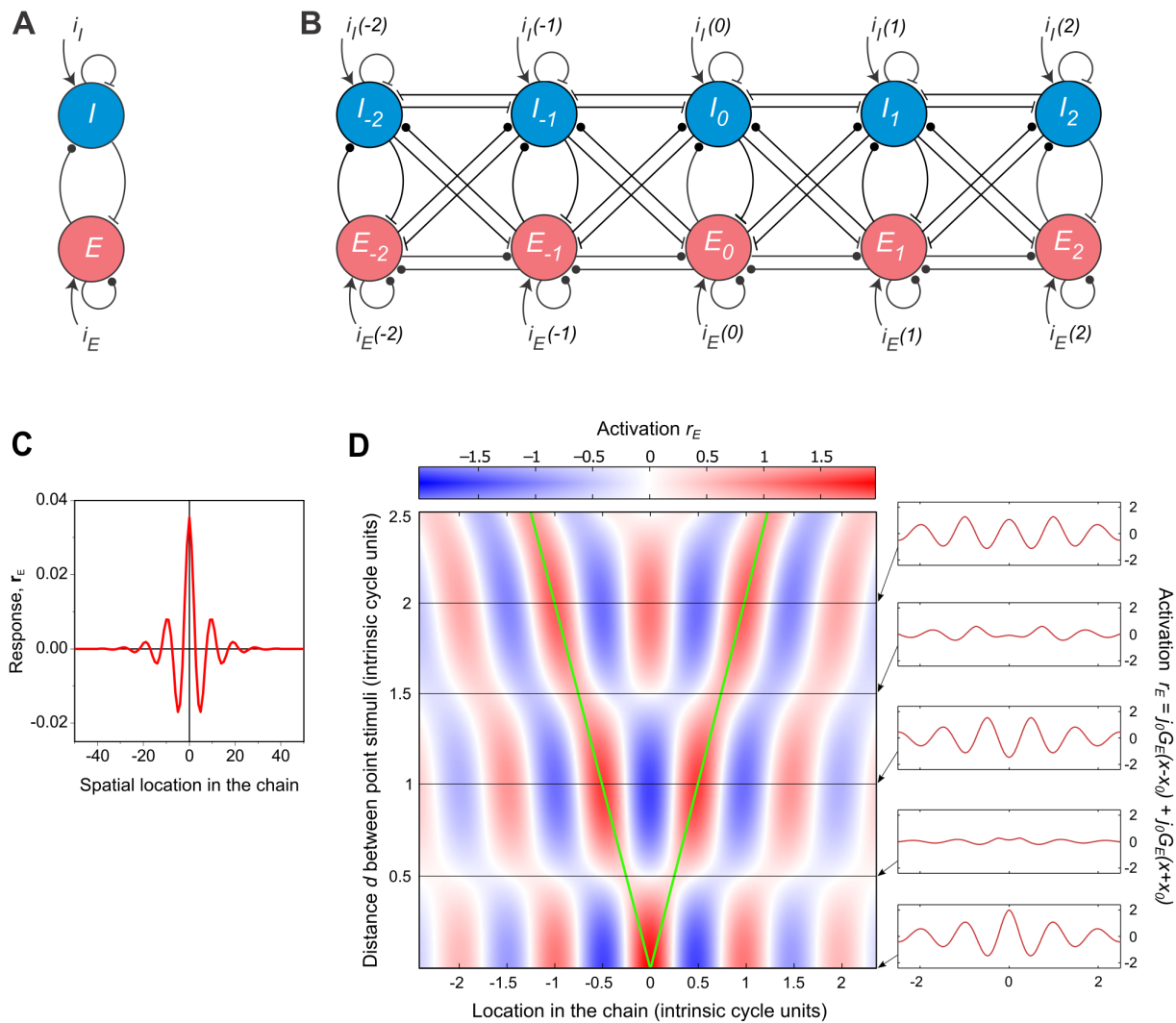

**Fig. 1. Distributed canonical circuit and its intrinsic tuning to spatial frequency.** (**A**) The generic Wilson-Cowan circuit contains one excitatory unit ($E$) and one inhibitory unit ($I$), connected reciprocally, each with recurrent feedback. The lines ending with circles and T-junctions represent, respectively, excitatory and inhibitory connections between the cells.

(**B**) The generic circuit is used as a motif in a distributed circuit. Indices $l$ of the motifs (notated as $E_l$ and $I_l$) indicate motif locations in the circuit. Currents $i_E(l)$ and $i_I(l)$ are the inputs into each motif generated by the stimulus.

(**C**) Response of the distributed circuit to a small "point" stimulus that activates a single motif in the circuit. The resulting point activation propagates through the circuit and forms a steady-state pattern: a neural wave that has the form of spatially damped oscillations. Only the excitatory component of the neural wave is shown.

(**D**) Interference of neural waves originating in two locations in the circuit, each activated by a point stimulus. The plot portrays a map of activation distributed across circuit locations (the abscissa) for multiple distances $d$ between two point stimuli (the ordinate). The activated locations are represented by green lines: starting at the same location at bottom ($d = 0$). The red and blue entries in the map represent positive and negative activation, explained in the color bar. The insets at right are horizontal sections through the map at five inter-stimulus distances (expressed in units of the intrinsic period of the circuit). As in panel C, only excitatory components of neural waves are shown.

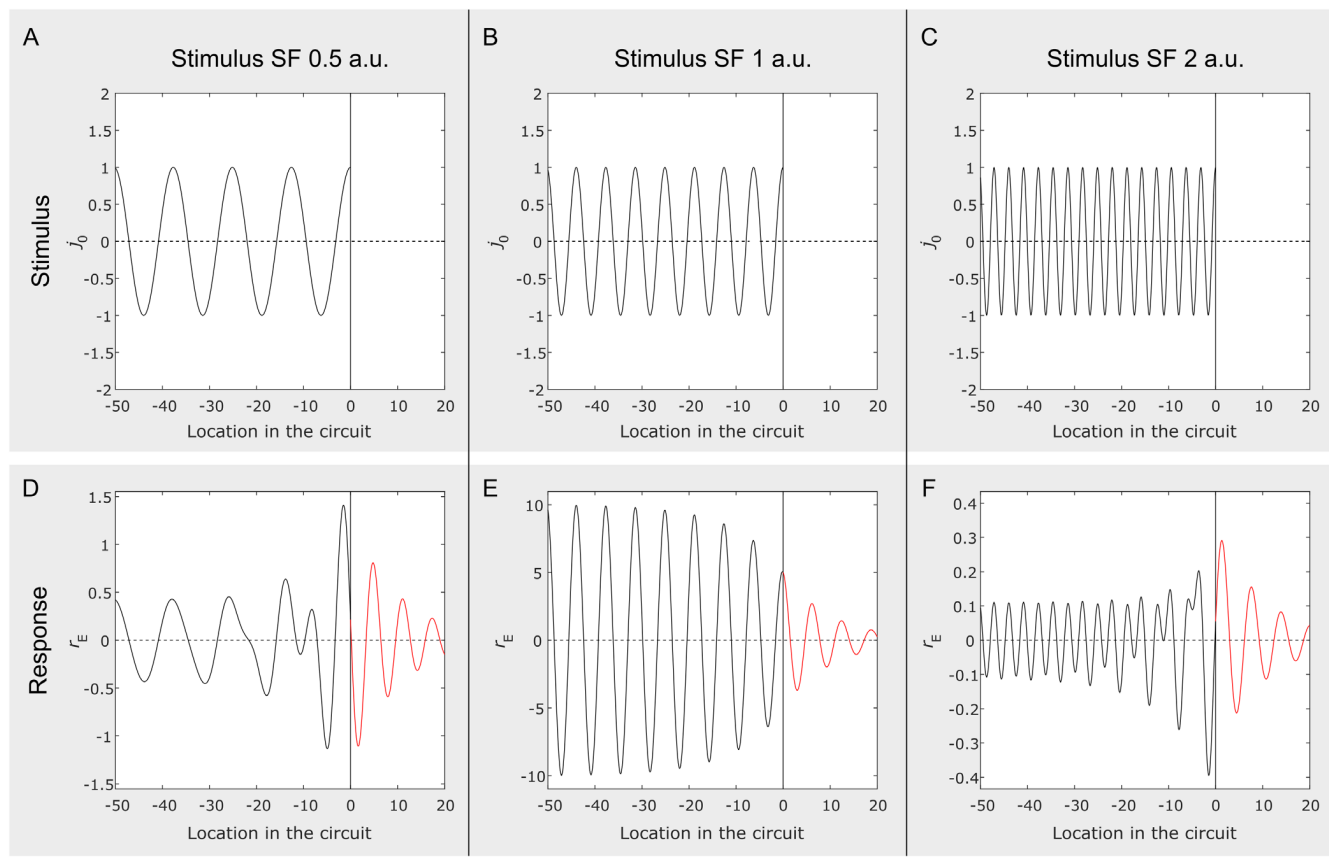

**Fig. 2. Response of the model circuit to luminance gratings. (A-C)** Three grating stimuli at the spatial frequencies of 0.5, 1 and 2 units of the intrinsic spatial frequency of the circuit. Stimulation is applied only to the part of the circuit that corresponds to negative values of the abscissa, with no direct stimulation for its positive values. **(D-F)** Simulated responses of the model circuit to stimuli in the same column: for circuit locations stimulated directly, in black, and for circuit locations outside of direct stimulation, in red. The responses were derived using Equation 4 in Methods. As in Figure 1C-D, only excitatory components of neural waves are shown.

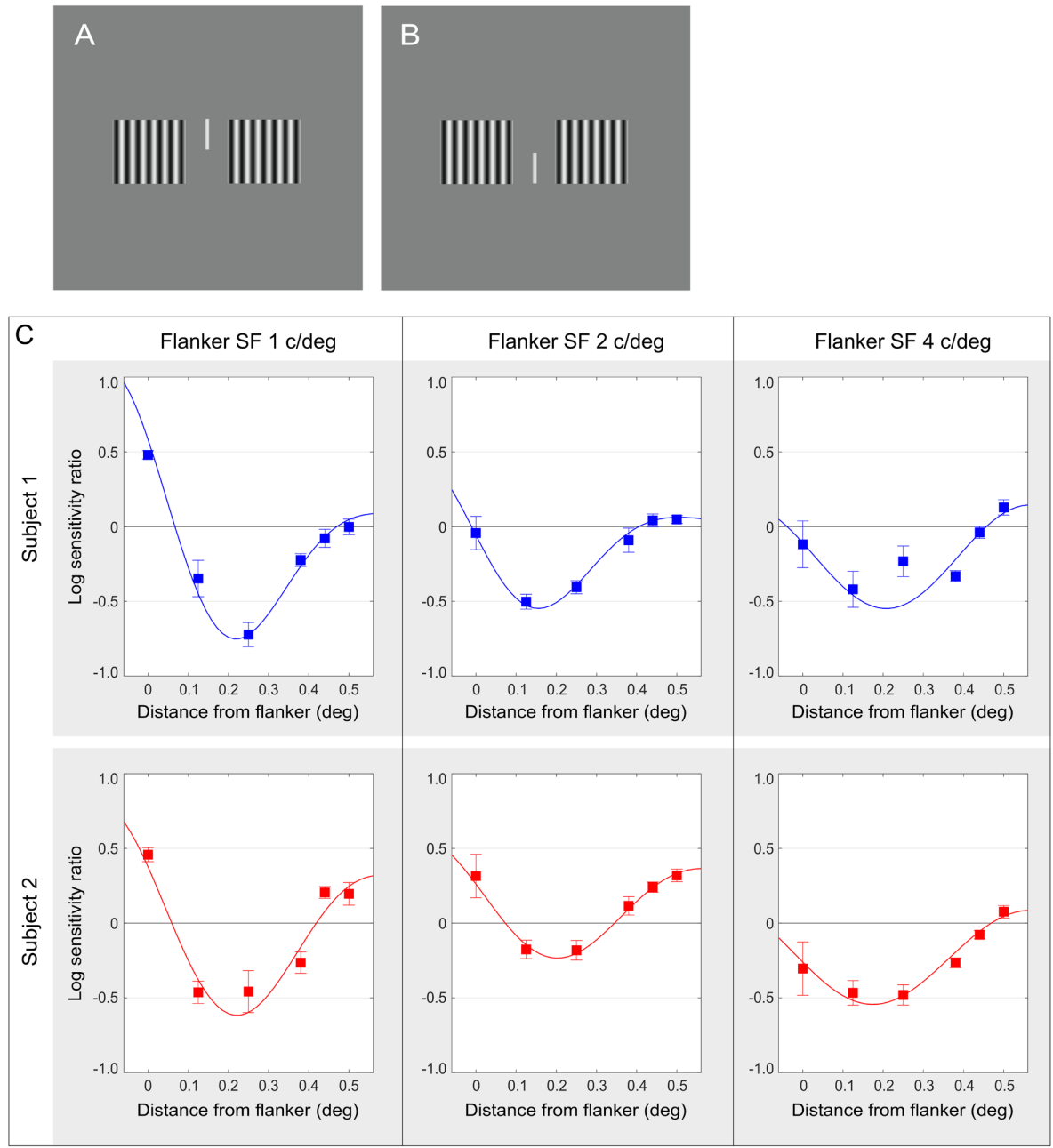

**Fig. 3. Measured contrast sensitivity in the region of lateral activation.** (**A-B**) The stimulus used in the psychophysical study. The probe was a faint vertical line (shown here as a high-contrast line) presented between two square patches of luminance grating ("flankers"). The probe appeared either above (A) or below (B) the horizontal midline. The task was to report whether the probe was seen in the upper or lower position. (**C**) In each panel, contrast sensitivity to the probe in the region of lateral activation is plotted as a function of distance from the flankers. (The data for probes presented left and right of the screen center were collapsed into a single function of distance from flanker edge.) Each column of plots contains data for a different spatial frequency (SF) of the flankers, and each row of plots contains data for a different human subject. The plots reveal a spatial waveform of lateral activation whose period does not systematically depend on the stimulus SF (see Supplemental Figure 1). These results are consistent with the model prediction that lateral activation is periodic and its SF is independent of the SF of direct stimulation.

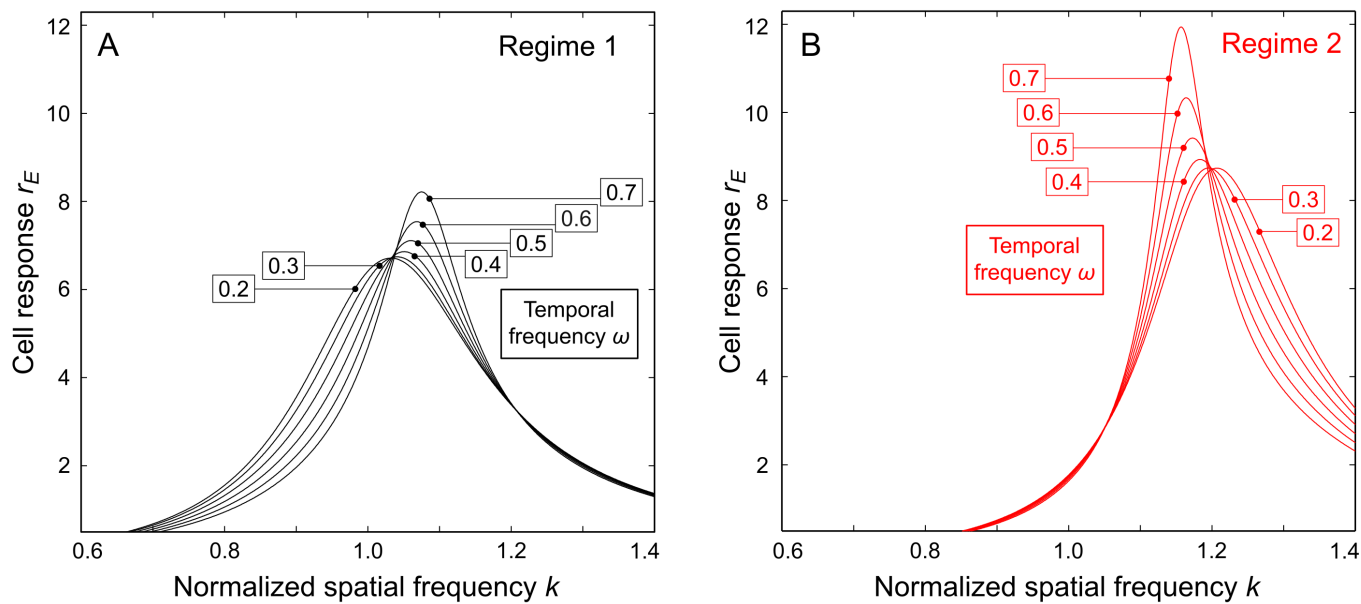

**Fig. 4. Maximum of model circuit response is predicted to depend on stimulus temporal frequency.** (**A**) Circuit response functions are plotted for different temporal frequencies (TF) of the stimulus in Regime 1. Function values are the amplitudes of neural waves computed using Equation 7 in Methods. The values of stimulus TF are displayed in the boxes. The maximum of each response function is obtained at the resonance spatial frequency (SFR) of the model circuit for the given stimulus TF. SFR increases with stimulus TF. (**B**) Response functions are plotted as in panel A for Regime 2. Here SFR decreases with stimulus TF.

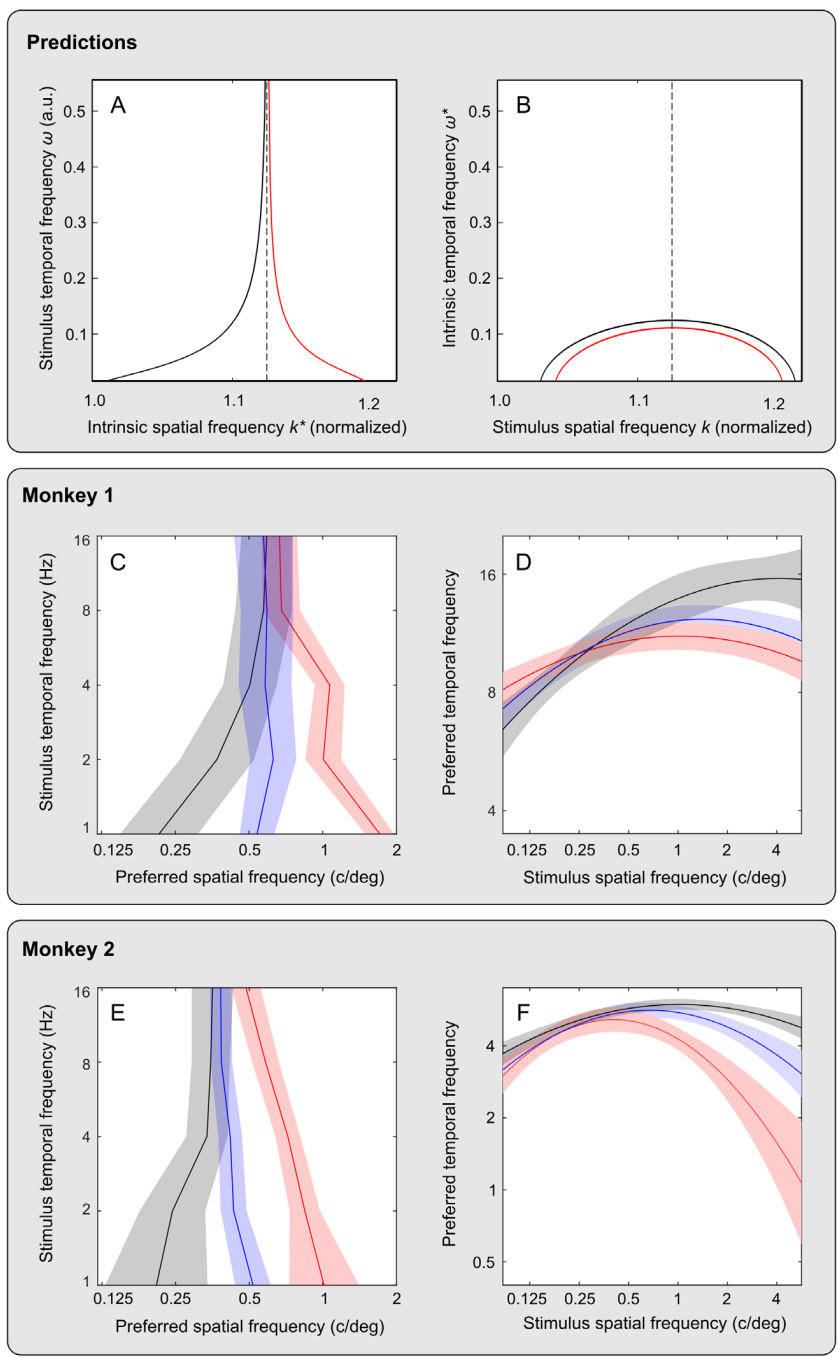

**Fig. 5. Spatial and temporal resonance in the model circuit and in macaque cortical neurons.** (**A-B**) Predictions of resonance spatial frequencies (SFR) for varying stimulus TF are displayed in A (summarizing maxima of response function in Figure 4), and predictions of resonance temporal frequencies (TFR) for varying stimulus SF are displayed in B (not shown Figure 4). The black and red curves represent two regimes of circuit activation. (**C-F**) Summaries of the preferred SF (C, E) and TF (D, F) measured in populations of neurons in area MT in two monkeys (Monkey 1 in C-D and Monkey 2 in E-F) for three stimulus contrasts. The lines and the shaded regions represent, respectively, the means and errors of the estimates: for low contrasts in black and gray, for medium contrasts in blue and light blue, and for high contrasts in red and pink. (See Figures 6-7 for detail of neuronal data and stimulus parameters.)

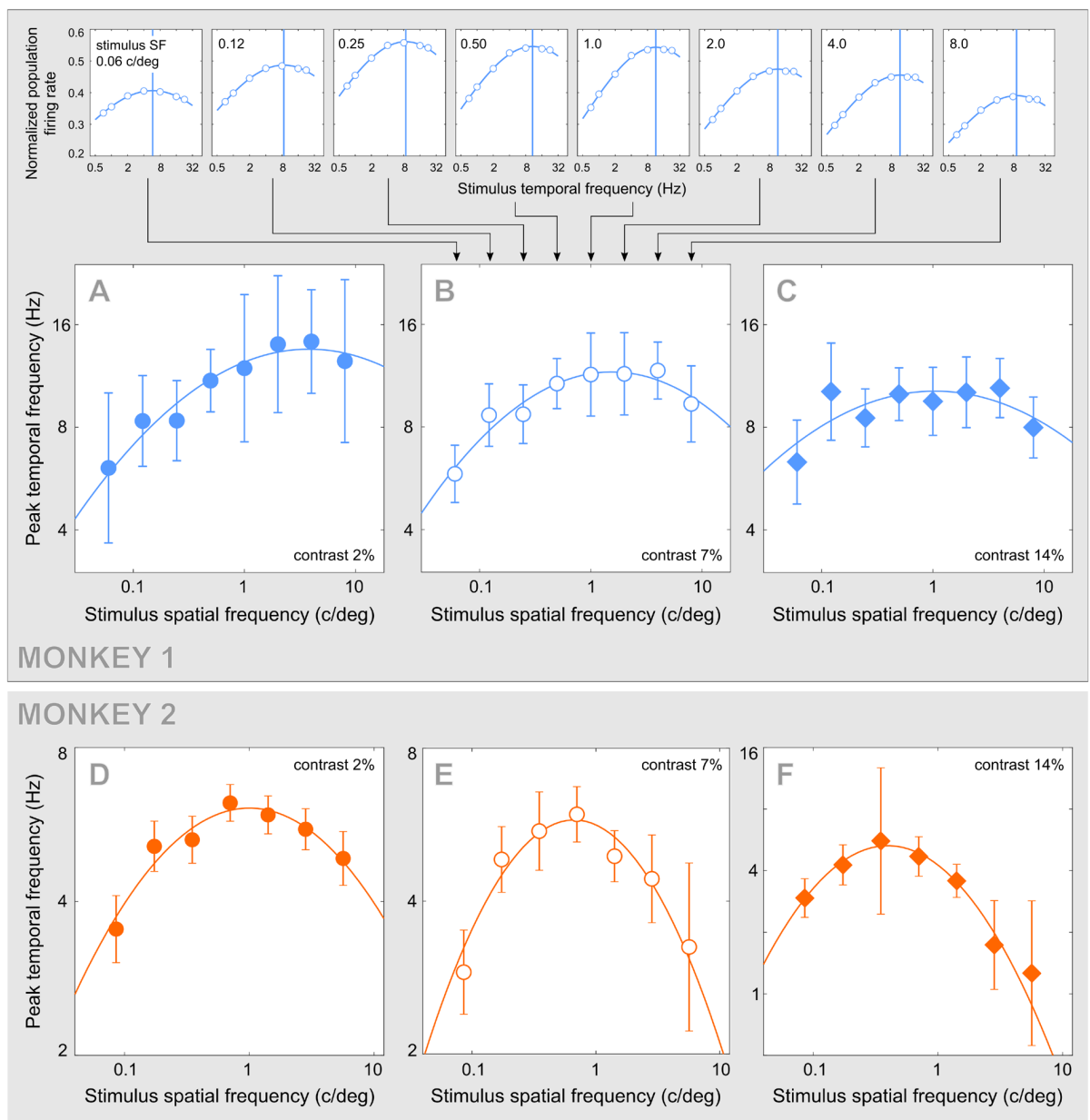

**Fig. 6. Preferred temporal frequency of MT neurons.** The top row of plots contains response functions measured in populations of MT neurons in Monkey 1 for eight stimulus SFs (marked at top left of each panel), all at the stimulus contrast of 7%. The blue vertical line in each panel marks the peak TF, assembled into a single function from eight stimulus SFs in panel B. (**A-C**) Summaries of peak TF estimated at three stimulus contrasts (marked at bottom right of each panel) for Monkey 1. The error bars are standard deviations of peak TF estimated by resampling (Methods). (**D-F**) Results of the same analysis for Monkey 2, in which peak TF could be estimated for in seven (rather than eight) conditions of stimulus SF. The shapes of these functions were as predicted by our model for contrasts 7% and 14% in Monkey 1 ($p=0.008$ and $p=0.004$, respectively; Methods), and for all contrasts in Monkey 2 ($p \ll 0.01$ for contrasts 2% and 7%, and $p=0.002$ for contrast 14%). Note that the ordinate range in panel F is enlarged to accommodate the wider range of peak TF in this contrast condition. Numbers of neurons used to perform the measurements are reported in Supplemental Figure 2.

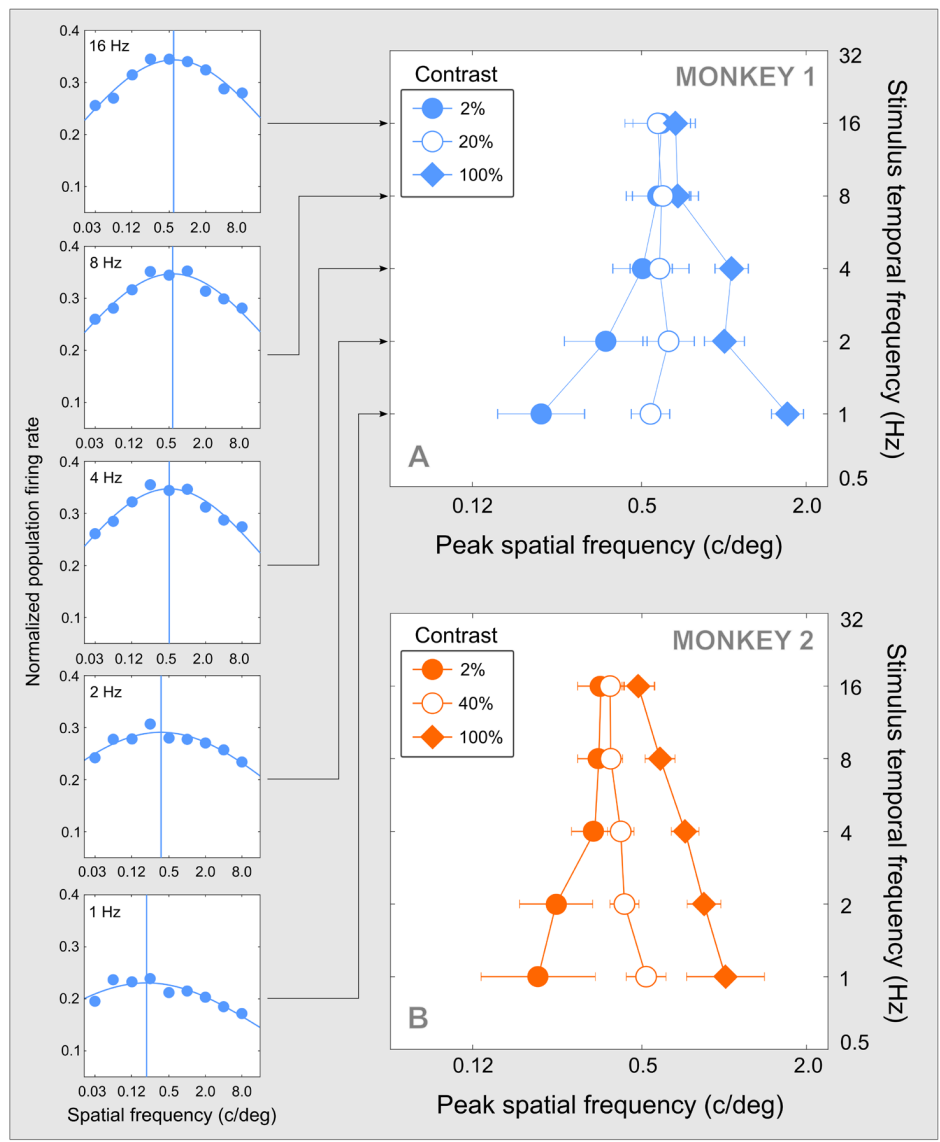

**Fig. 7. Preferred spatial frequency of MT neurons.** (**A**) The left column of plots contains response functions measured in MT neurons of Monkey 1 at the same stimulus contrast (2%) for five stimulus TFs (labelled at top left of each panel). The blue vertical line in each panels marks an estimate of peak SF. The five estimates are assembled into a single function at left (filled circles) in the larger graph. This graph contains two other functions of peak SF estimated in the same manner for stimulus contrasts of 20% and 100% in the same monkey.

(**B**) Results of the same analysis for Monkey 2 are shown in panel B. The results displayed for two monkeys were obtained in the same stimulus conditions with the only difference that medium contrasts were 20% for Monkey 1 and 40% for Monkey 2. (Medium contrasts were chosen so that their plots occupied intermediate positions between the low- and high-contrast plots.) The error bars are standard deviations of peak SF estimated by resampling (Methods). Between stimulus contrasts of 2% and 100%, peak SF increased significantly in both monkeys: for TFs below 8 Hz in Monkey 1 ($p<0.01$; Methods) and for all TFs in Monkey 2 ($p<0.03$). Within contrast, peak SF increased with TF at the contrast of 2% ($p<0.01$ in Monkey 1; $p=0.05$ in Monkey 2) and decreased with TF at the contrast of 100% ($p<0.01$ in Monkey 1 and $p=0.02$ in Monkey 2). Numbers of neurons used to perform the measurements are reported in Supplemental Figure 3.

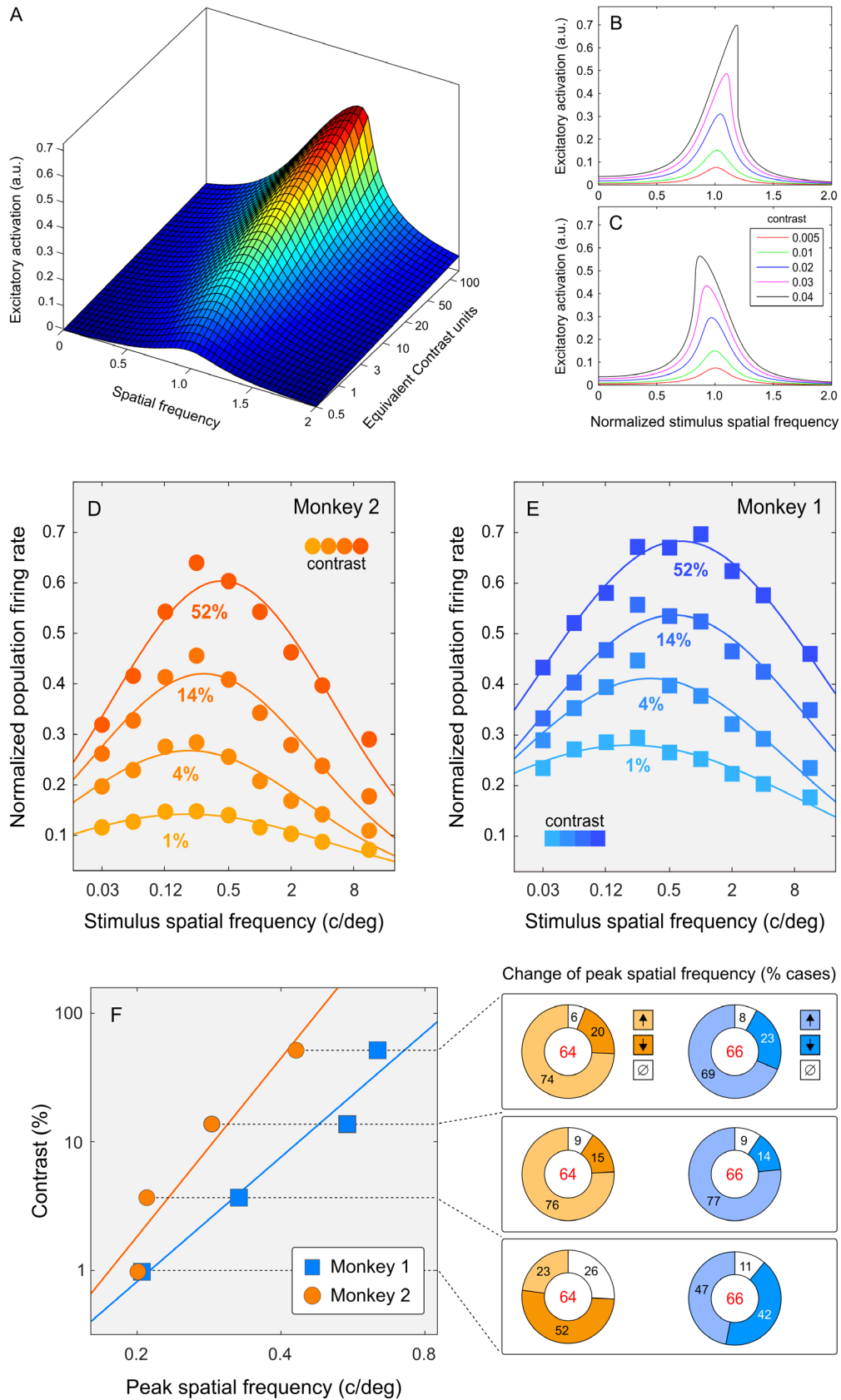

**Fig. 8. Response of the model circuit and cortical neurons to increasing stimulus contrast.** (**A**) Theoretical response surface of the circuit obtained by iterating Equation 15 in Methods.

(**B-C**) The curves are slices through the surface in panel A. Each curve represents the amplitude of the excitatory component of neural wave in the model circuit at one stimulus contrast, plotted as a function of stimulus SF. Stimulus SF is normalized to the resonance SF (SFR) of the circuit measured at the lowest tested contrast (the red curve). Increasing stimulus contrast can increase SFR, as in panels A-B, or decrease it, as in panel C (Equation 17 in Methods).

(**D-E**) Spatial response functions measured in populations of MT neurons are plotted separately for four stimulus contrasts (separated by color) for Monkey 1 in panel E and Monkey 2 in panel D. The data are collapsed across TF.

(**F**) Peak SFs of neural populations are plotted for four stimulus contrasts, summarizing results in panels E (Monkey 1) and D (Monkey 2), using colors matched by monkey. In both monkeys, peak SF increases with contrast in agreement with model predictions. The donut charts at right display percentages of cases in which peak SF significantly ($p < 0.05$) increased (↑), decreased (↓), or did not change (∅) for each step of stimulus contrast. (Black numerals represent percentages and red numerals represent numbers of neurons used in the analysis.) Significance was established using the Wilcoxon rank-sum test.

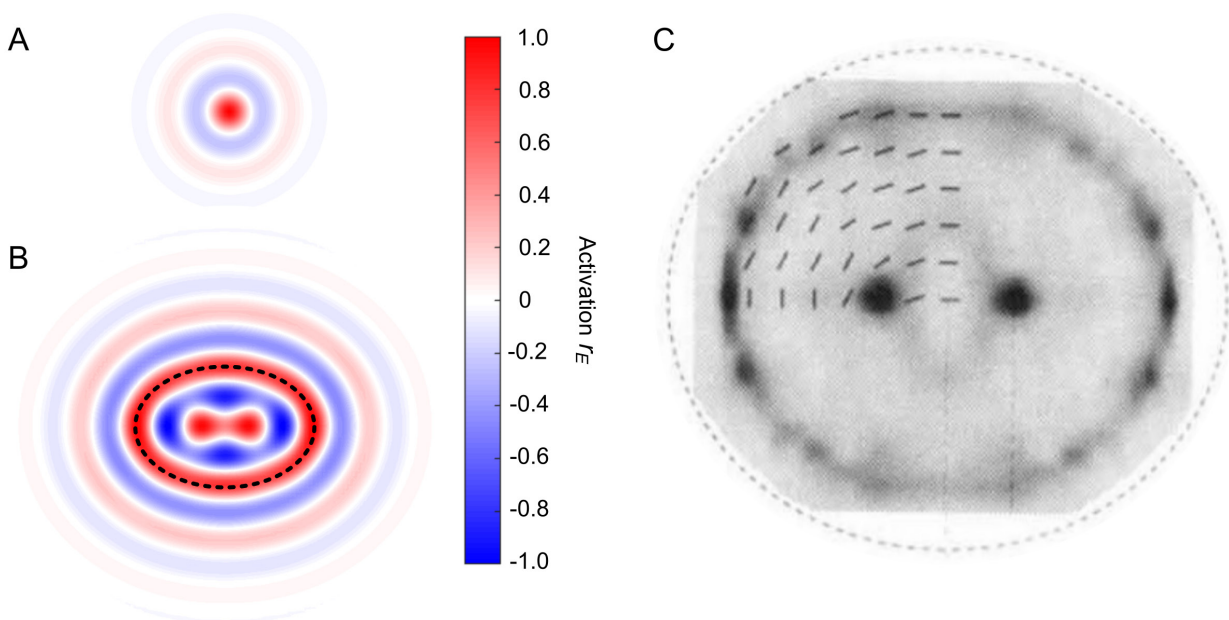

**Fig. 9. Interference of neural waves in two spatial dimensions.** (**A**) Response of the two-dimensional excitatory-inhibitory circuit to a "point" stimulus (*cf.* Figure 1C). Colors represent circuit activation $r_E$ explained in the color bar. The same color bar applies to panels A-B. (**B**) Response of the model circuit to an elliptical ring stimulus (represented by the dashed black contour). Note two regions of positive activation near the focal points of the stimulus ring, predicting that contrast threshold should reduce in these regions as compared to other regions inside the ring. (**D**) Results of a psychophysical study in which a high-contrast elliptical ring was used as the stimulus (*26*). Gray levels represent contrast thresholds measured at multiple locations inside the ring. Dark colors represent reduction of threshold caused by the stimulus. Note that the threshold was reduced near the focal points of the ellipse, represented by two dark regions, in agreement with model predictions in B.

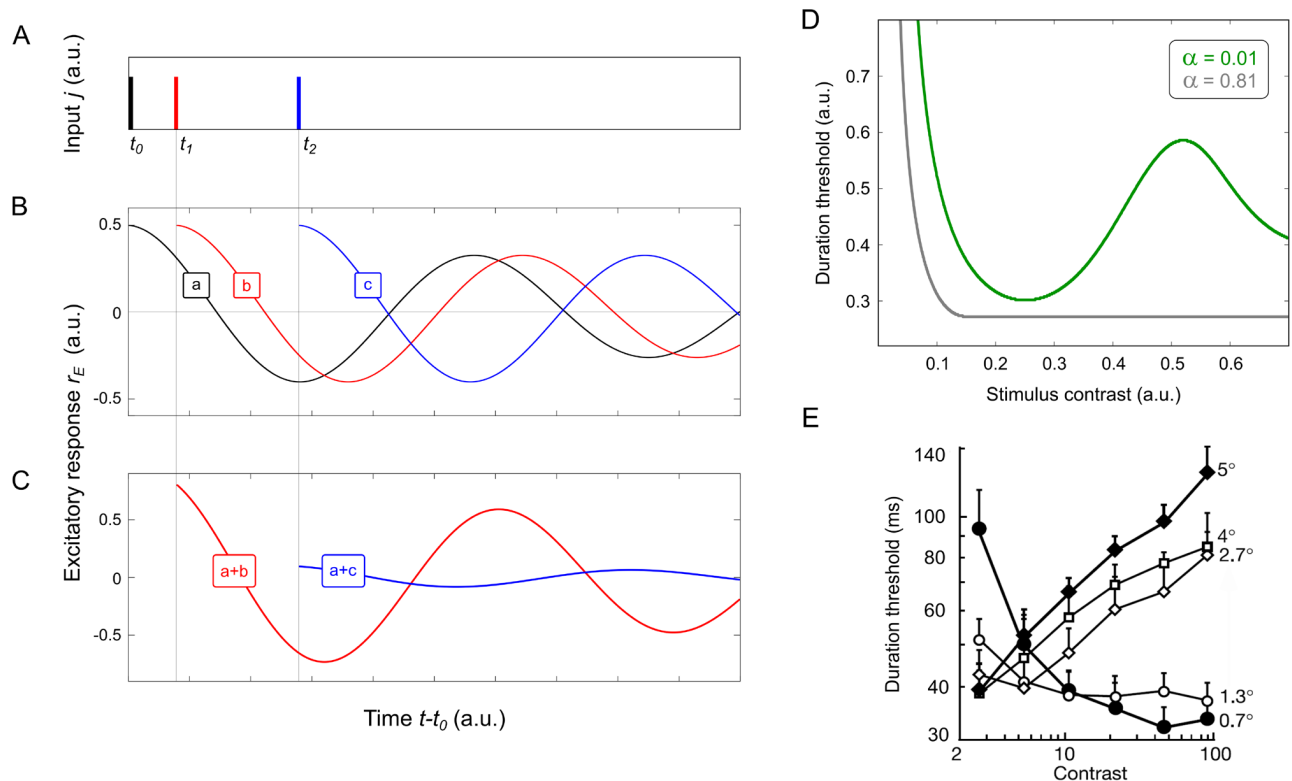

**Fig. 10. Interference of temporal oscillations and duration threshold.** (**A-C**) Temporal oscillation 'a', 'b' and 'c' (shown in B) are elicited by impulses of brief stimulation at instants $t_0$, $t_1$ and $t_2$ (shown in A). The oscillations interfere with one another across time, forming patterns of activity of which two examples are shown in C, obtained by superposition of 'a' and 'b' ('a+b') and superposition of 'a' and 'c' ('a+c'). Such interference patterns add up to higher or lower activity, making the system respectively more or less sensitive to visual stimulation, during stimulation or after the stimulus is turned off.

(**D**) By simulating temporal interference of oscillations in our single-node model (Equation 5 in Supplemental Modeling Methods), we found that stimulus visibility is expected to depend on stimulus duration and luminance contrast. Effects of these two parameters of stimulation are predicted to interact, leading to counterintuitive results. Here we show how stimulus contrast is predicted to affect *duration threshold*, which is the stimulus duration at which the stimulus is just visible. When the circuit is dominated by excitation (model parameter $\alpha$ is high; Section 1.3.1 in Methods), duration threshold is predicted to monotonically decline with stimulus contrast, represented by the gray curve. When the circuit is dominated by inhibition ($\alpha$ is low), duration threshold is predicted to form a non-monotonic function of stimulus contrast, represented by the green curve. Here, the model predicts that threshold first declines with contrast, but further increase of contrast is expected to raise threshold, before it declines again.

(**E**) Psychophysical studies in human subjects (*47-48, 52*) found that duration threshold can increase or decrease with stimulus luminance contrast depending on stimulus size (indicated in the figure to the right of each curve). Since stimulus size is known to determine the amount of inhibition in cortical circuits (*35-36*), our model can explain the reversal of relationship between stimulus contrast and duration threshold shown in E (*47*): from threshold decreasing with contrast for small stimuli (corresponding to the gray curve in D) to threshold increasing with contrast for large stimuli (green curve in D).

# 1   Methods

## 1.1   Physiological methods

**Animals.**   Two adult male rhesus monkeys (*Macaca mulatta*) of ages 11 and 12 were used in this study. Experimental protocols were approved by the Salk Institute Animal Care and Use Committee, and conform to US Department of Agriculture regulations and to the National Institutes of Health guidelines for the humane care and use of laboratory animals. Procedures for surgery and wound maintenance have been described in detail elsewhere (Dobkins and Albright, 1994; Pawar et al., 2019).

**Apparatus.**   All visual stimuli were generated using Matlab (Mathworks, Natick) software using a high-resolution graphics display controller (Quadro Pro Graphics card, 1024×768 pixels, 8 bits/pixel) operating in a Pentium class computer. Stimuli were displayed on a 21-inch monitor (75 Hz, non-interlaced, 1024×768 pixels; model GDM-2000TC; Sony, Tokyo, Japan). The output of the video monitor was measured with a PR650 photometer (Photo-Research, Chatsworth, CA), and the voltage/luminance relationship was linearized independently for each of the three guns in the cathode ray tube.

**Behavioral procedure.**   Monkeys were seated in a standard primate chair (Crist Instruments, Germantown, MD) with the head post rigidly supported by the chair frame. The task was to fixate a small (0.2° diameter) fixation target in the presence of moving visual stimuli for the duration of each trial (500-2000 msec). The target was presented on a video display at a viewing distance of 57 cm in a dark room ($<0.5$ cd/m$^2$). The mean background luminance of the monitor was 15 cd/m$^2$. Eye position was sampled at 120 Hz using an infrared video-based system (IScan, Burlington, MA). The eye position data were monitored and recorded with the CORTEX program (Laboratory of Neuropsychology, National Institute of Mental Health, Bethesda, MD), which was also used to implement the behavioral paradigm and to control stimulus presentation. After eye position was maintained within a 2° window centered on the fixation target throughout the trial, animals were given a small (0.15 cc) juice reward.

**Electrophysiological procedure.**   A craniotomy was performed to allow for electrode passage into area MT. Activity of single units was recorded in area MT using tungsten microelectrodes (3–5 MΩ; Frederick Haer Company, Bowdoinham, ME), which were driven into cortex using a hydraulic micropositioner (model 650; David Kopf Instruments, Tujunga, CA). Neurophysiological signals were filtered, sorted, and stored using the Plexon (Dallas, TX) system. Visual responses were recorded from 139 directionally selective MT neurons in two awake, fixating macaque monkeys (74 and 65 neurons in Monkeys 1 and 2 respectively). We measured firing rates to stimuli at five to seven different levels of luminance contrast (0.05-100%) at the preferred spatiotemporal frequencies: five SFs and one to five TFs. The different stimulus conditions and contrasts were interleaved in random order across trials.

**Data resampling.**   For each neuron, the firing rates estimated in separate trials within each condition of stimulus frequency and contrast were resampled with replacement. The number of samples was ten (which is the number of trials employed in the experiments). Response functions were fitted to the resampled data using non-parametric polynomial regression, repeated for 500 iterations of resampling to estimate errors of peak SF within each condition. The errors were used to measure differences between peaks across stimulus contrasts. A similar procedure was used to estimate errors of peak TF for each condition.

**Data analysis for Figure 6.** Each point in the top plots of Figure 6 was obtained by averaging firing rates within a sliding kernel defined on three stimulus dimensions: contrast, SF, and TF. For example, in the top left plot, the kernel contained four values of stimulus SF (geometric mean 0.06 c/°, indicated in the plot), three values of stimulus contrast (yielding the geometric mean of 7%), and one value of stimulus TF. The kernel was then advanced by one TF step while encompassing the same range of values of stimulus SF and contrast. Preferred values of stimulus TF were then estimated (marked by vertical lines in each top panel) and assembled into a temporal preference function (panel B). Three such preference functions are displayed in Panels A–C for stimulus contrasts of 2, 7 and 14%, using Monkey 1 data. The same analysis was performed using Monkey 2 data (panels D–F) as explained in the caption of Figure 6.

**Data analysis for Figure 7.** Similar to Figure 6, a sliding three-dimensional kernel was used to average firing rates. For example, at the first point in the top left panel, the firing rate was measured for the kernel containing three values of stimulus TF (geometric mean 16 Hz, indicated in the plot), two values of stimulus contrast (geometric mean 2%), and three values of stimulus SF (geometric mean 0.03 c/°). The kernel was then advanced by one SF step while encompassing the same range of values of stimulus TF and contrast. Preferred values of stimulus SF were then estimated (marked by vertical lines in each panel at left) and assembled into a spatial preference function. Such preference functions are plotted in panel A for stimulus contrasts of 2, 20 and 100%, using Monkey 1 data. The same analysis was performed using Monkey 2 data for stimulus contrasts of 2, 40 and 100% (panel B) as explained in the caption of Figure 7.

## 1.2   Psychophysical methods

**Subjects.** Two adult subjects with normal or corrected-to-normal vision took part in the study. The subjects were given sufficient practice with the experimental task prior to the experiment.

**Apparatus.** The experiments were carried out in a dark room ($< 0.5$ cd/m$^2$). The stimuli were presented on a 21-inch monitor (SONY Color Graphic Monitor GDM-500PS) under the control of a personal computer with a commercially available circuit that provided high gray-scale resolution of 14 bits (Li at al, 2003). Visual stimuli were generated by custom software and presented using the commercial software package Matlab (Mathworks, Natick). The monitor was gamma calibrated and had a resolution of 1600×1200 pixels (horizontal×vertical) with the vertical frame rate of 160 Hz. Subjects viewed stimuli binocularly from a distance of 57 cm using a chin-and-head rest.

**Stimulus.** The stimulus consisted of two square patches of vertical stationary luminance grating ("flankers") and a vertical line ("probe") positioned between the flankers. The mean luminance of the screen and of the flankers was 24.65 cd/m$^2$. The flankers had one of five contrasts (10, 30, 50, 70 or 90%) and one of three spatial frequencies (1, 2, or 4 c/°). Inner edges of flankers were 0.5° away from the screen center. The probe was a faint vertical line that appeared above or below the horizontal midline of the screen (as shown in panels A and B of Figure 3) at one of five distances (0, 0.125, 0.25, 0.375, 0.4375, or 0.5°, indicated in the abcissa of Figure 3C) from one of the inner edges of the flankers. Probe contrast was controlled by an adaptive staircase procedure described just below. The task was to report whether the probe was seen in the upper or lower position.

**Procedure.** Each trial started with a fixation dot, which had a Gaussian luminance profile with the spatial constant of $0.17°$, presented at screen center for 200 ms. Fixation was followed by the stimulus presented for 250 ms. Subjects used the upward and downward arrow keys on the the keyboard to report whether they saw the probe appear in the upper or lower position. Subject's response triggered the next trial. Contrast detection threshold of the probe was measured for each stimulus condition (spatial distance from the flanker, flanker spatial frequency, and flanker contrast) using a 3-down, 1-up adaptive staircase procedure. Starting with an initial contrast of 40%, probe contrast was either reduced, following three correct responses, or increased, following a single incorrect response. Contrast steps (same up and down) were 12% up to the second reversal of contrast, after which contrast steps were reduced to 3%. The procedure was terminated after 30 trials. Contrast threshold was estimated by averaging the last five values of contrast selected by the procedure in each stimulus condition. Experimental runs with the same stimulus conditions were repeated in each subject three-to-four times. Since we found no statistically significant differences between estimates of probe detection threshold at different flanker contrasts, the data presented in Figure 3C are averaged across flanker contrasts.

## 1.3 Modeling methods

### 1.3.1 Definition of the distributed E/I network

We have studied a system of equations for a chain of Wilson-Cowan motifs (Wilson and Cowan, 1972) each containing an excitatory cell ("$E$") and an inhibitory cell ("$I$") as shown in Figure 1A. The generic form of the system is

$$\tau_E \frac{dr_E(l)}{dt} = -r_E + g_E(\mathcal{W}_E),$$
$$\frac{dr_I(l)}{dt} = -r_I + g_I(\mathcal{W}_I), \tag{1}$$

where $r_E$ and $r_I$ represent the firing rates of the excitatory and inhibitory cells, $\tau_E$ represents the relaxation time of excitation (in units of the relaxation time of inhibition), $l$ is the location index in the chain, $g_E$, $g_I$ are sigmoid functions. The terms $\mathcal{W}_E$ and $\mathcal{W}_I$ represent sources of cell activation (from stimulation and from other cells in the network) at location $l$:

$$\mathcal{W}_E = \Big[ w_{EE} r_E(l) + \tilde{w}_{EE} r_E(l+1) + \tilde{w}_{EE} r_E(l-1) \Big] - \Big[ w_{EI} r_I(l) + \tilde{w}_{EI} r_I(l+1) + \tilde{w}_{EI} r_I(l-1) \Big] + i_E(l, t),$$

$$\mathcal{W}_I = \Big[ w_{IE} r_E(l) + \tilde{w}_{IE} r_E(l+1) + \tilde{w}_{IE} r_E(l-1) \Big] - \Big[ w_{II} r_I(l) + \tilde{w}_{II} r_I(l+1) + \tilde{w}_{II} r_I(l-1) \Big] + i_I(l, t),$$

where $w$ and $\tilde{w}$ represent the weights of connections respectively within and between the motifs (Savel'ev and Gepshtein, 2014).

When modulations of neural activity occur on a spatial scale significantly larger than the distance between the nearest network motifs, and at stimulus contrasts for which the system response is far below saturation, the excitatory/inhibitory network can be modeled by a system of nonlinear partial differential equations (derived in Supplementary Materials) of the form:

$$\frac{\partial r_E}{\partial t} = -r_E + W_{EE} r_E + D_{EE} \frac{\partial^2 r_E}{\partial x^2} - W_{EI} r_I - D_{EI} \frac{\partial^2 r_I}{\partial x^2} + \alpha j - \gamma_E \left( \frac{\partial r_E}{\partial t} + r_E \right)^3,$$

$$\frac{\partial r_I}{\partial t} = -r_I + W_{IE} r_E + D_{IE} \frac{\partial^2 r_E}{\partial x^2} - W_{II} r_I - D_{II} \frac{\partial^2 r_I}{\partial x^2} + (1 - \alpha) j - \gamma_I \left( \frac{\partial r_I}{\partial t} + r_I \right)^3, \tag{2}$$

where $r_E(x, t)$ and $r_I(x, t)$ are firing rates of excitatory and inhibitory neurons at position $x$ and time $t$, and where $\gamma_E$ and $\gamma_I$ are Taylor expansion coefficients that arise due to non-linearity of Wilson-Cowan sigmoid activation function (Wilson and Cowan, 1972, p. 4). The weights $W_{EE}$, $W_{EI}$, $W_{IE}$, $W_{II}$ control local interactions:

$$W_{EE} = w_{EE} + 2\tilde{w}_{EE}, \ W_{EI} = w_{EI} + 2\tilde{w}_{EI}, \ \text{etc.}$$

And the terms $D_{EE}$, $D_{EI}$, $D_{IE}$, and $D_{II}$ are responsible for spatial spread of activation:

$$D_{EE} = \tilde{w}_{EE}, \ D_{EI} = \tilde{w}_{EI}, \ \text{etc.}$$

The stimulus is represented by the input current $j(x,t)$ divided between the excitatory and inhibitory neurons with fractions $\alpha$ and $1-\alpha$, such that the excitatory input is $\alpha j$ and the inhibitory input is $(1-\alpha)j$.

### 1.3.2 Intrinsic wave

**Point activation.** First, we considered the system responds to a "point" stimulus $j(x,t) = \delta(x)$, where $\delta(x)$ is the Dirac delta function. The stimulus activates a singe motif of the chain. In linear approximation, the static solution has the form:

$$
\begin{aligned}
G_E &= e^{-\lambda|x|}\left(\Gamma_E \cos(k_n x) - \Delta_E \text{sign}(x)\sin(k_n x)\right), \\
G_I &= e^{-\lambda|x|}\left(\Gamma_I \cos(k_n x) - \Delta_I \text{sign}(x)\sin(k_n x)\right),
\end{aligned}
\tag{3}
$$

where $\text{sign}(x) = x/|x|$ for $x \neq 0$ and $\text{sign}(0) = 0$. Equation 3 determines a spatial oscillation (Figure 1C). The intrinsic spatial frequency $k_n$, the rate of spatial decay $\lambda$, and parameters that determine the amplitude of oscillation ($\Gamma_E$, $\Gamma_I$, $\Delta_E$, $\Delta_I$) are functions of the weights of neuronal connections $W_{EE}$, $W_{EI}$, $W_{IE}$, $W_{II}$ and $D_{EE}$, $D_{EI}$, $D_{IE}$, $D_{II}$ (Savel'ev and Gepshtein, 2014).

**Distributed activation.** For stimuli more complex than a delta function, the general static solution can be written in linear approximation as:

$$
\begin{aligned}
r_E &= \int_{-\infty}^{\infty} j(x')G_E(x-x')dx', \\
r_I &= \int_{-\infty}^{\infty} j(x')G_I(x-x')dx',
\end{aligned}
\tag{4}
$$

suggesting how our model is related to the standard description of visual neural mechanisms in terms of linear filters. For example, for two point stumuli described by $j = \delta(x - l/2) + \delta(x + l/2)$ separated by distance $l$, the excitatory response is defined simply as $r_E = G_E(x-l/2) + G_E(x+l/2)$ (Figure 1D).

### 1.3.3 Linear resonance analysis

To understand the interaction of systems responses to stimulus spatial ($k$) and temporal ($\omega$) frequencies, we studied system response to input current $j(x,t)$ produced by drifting luminance gratings $j = j_0 \cos(kx - \omega t)$ as stimuli. In linear approximation, the response of system (2) to such spatiotemporally harmonic stimuli is

$$
\begin{aligned}
r_E(x,t) &= \mathcal{E}_C \cos(kx - \omega t) + \mathcal{E}_S \sin(kx - \omega t), \\
r_I(x,t) &= \mathcal{I}_C \cos(kx - \omega t) + \mathcal{I}_S \sin(kx - \omega t).
\end{aligned}
\tag{5}
$$

By substituting (5) in (2), we obtained algebraic equations for amplitudes $\mathcal{E}_C, \mathcal{E}_S, \mathcal{I}_C$, and $\mathcal{I}_S$. As in our nonlinear analysis of resonance behavior below (13), the algebraic equations can be written in matrix form:

$$\mathcal{H}\mathbf{Z} = \mathbf{I}_0, \tag{6}$$

where

$$\mathcal{H} = \begin{bmatrix} W_{EE} - 1 - k^2 D_{EE} & \omega & -W_{EI} + k^2 D_{EI} & 0 \\ -\omega & W_{EE} - 1 - k^2 D_{EE} & 0 & -W_{EI} + k^2 D_{EI} \\ W_{IE} - k^2 D_{IE} & 0 & -W_{II} - 1 + k^2 D_{II} & \omega \\ 0 & W_{IE} - k^2 D_{IE} & -\omega & -W_{II} - 1 + k^2 D_{II} \end{bmatrix}$$

is the matrix of wave-component interaction, $\mathbf{Z} = (\mathcal{E}_C, \mathcal{E}_S, \mathcal{I}_C, \mathcal{I}_S)^T$ is the wave-component vector, and $\mathbf{I}_0 = -j_0(\alpha, 0, 1 - \alpha, 0)^T$ is the stimulation.

Elements of the matrix $\mathcal{H}$ describe interaction of neural wave components. These elements depend on spatial and temporal stimulus frequencies, $k$ and $\omega$, as well as on coefficients of interaction between cells: local coefficients $W$ and propagation coefficients $D$. In effect, spatial resonance depends on $\omega$ and temporal resonance depends on $k$, as we show below.

Using Cramer's rule, the solution of (6) can be written as

$$\mathbf{Z} = (\det \mathcal{H}_1, \det \mathcal{H}_2, \det \mathcal{H}_3, \det \mathcal{H}_4) / \det \mathcal{H}, \tag{7}$$

where $\mathcal{H}_i$ are standard matrices derived from $\mathcal{H}$ by substituting its corresponding columns with $\mathbf{I}_0$. We used (7) to derive Figure 4.

To estimate stimulus frequencies at which the firing rate reaches its maximum (thus defining system selectivity), we assumed that the denominator $H = \det \mathcal{H}$ of the solution for $\mathbf{Z}$ has a corresponding minimum. In this approximation, the problem of finding the maximum of activity is reduced to the analysis of $H$, which has this form:

$$H = \underbrace{\mu \left[ (k^2 - k_n^2 + \lambda^2)^2 + 4k_n^2 \lambda^2 \right]^2}_{\text{Term A}} + \underbrace{\omega^2 \left[ \kappa_4 k^4 - \kappa_2 k^2 \right]}_{\text{Term B}} + \underbrace{\omega^2 \left[ \omega^2 + \kappa_0 \right]}_{\text{Term C}}, \tag{8}$$

where $\mu, \kappa_4, \kappa_2, \kappa_0$ are weight-dependent constants. Note that Term A depends on $k$ and not on $\omega$, whereas Term C depends on $\omega$ and not on $k$. Accordingly, Term A describes the system's spatial selectivity (at $\omega = 0$), Term C describes the system's temporal selectivity (at $k = 0$), and Term B, which contains products of $\omega$ and $k$, describes the interaction of temporal and spatial stimulus dimensions.

We used (8) to derive the spatial (Figure 5A) and temporal (Figure 5B) maxima of the system response by finding the minima of $H$, respectively at fixed $\omega$ and fixed $k$. In other words, the condition $\partial H / \partial k = 0$ is satisfied at the resonance spatial frequency $k_r(\omega)$, and the condition $\partial H / \partial \omega = 0$ is satisfied at the resonance temporal frequency $\omega_r(k)$. As a result, the frequency of spatial resonance is given (implicitly) by

$$\omega^2 = [(k_r^2 - k_n^2 + \lambda^2)^2 + 4k_n^2 \lambda^2](k_r^2 - k_n^2 + \lambda^2) / (\mu_1 - \mu_2 (k_r^2 - k_n^2 + \lambda^2)), \tag{9}$$

where $\mu_1$ and $\mu_2$ are weight-dependent constants, and the sign of the denominator determines whether $k_r$ increases or decreases with $\omega$. The vertical asymptote (displayed in Figure 5A-B) that separates two regime of system behavior is given by the conditions in which the denominator of (9) is zero, i.e., at

$$k_r^2 = k_n^2 - \lambda^2 + \frac{\mu_1}{\mu_2}.$$

And the resulting condition of temporal resonance is given by

$$\omega_r^2 = \frac{1}{2} (\kappa_2 k^2 - \kappa_4 k^4 - \kappa_0). \tag{10}$$

### 1.3.4   Nonlinear resonance analysis

To study circuit behavior beyond linear approximation, and to investigate how resonance frequency of the circuit should change with contrast, we derived system response to spatially periodic stimuli of the form $j(x) = j_0 \cos kx$, where $k$ is spatial frequency. By substituting into (2) an approximate solution in the form

$$r_E(x) = \mathcal{E} \cos kx, \tag{11}$$

$$r_I(x) = \mathcal{I} \cos kx, \tag{12}$$

while ignoring higher order spatial harmonics (such as $\cos 3kx$ and $\sin 3kx$), we obtained a nonlinear algebraic equation for the amplitudes of excitatory ($\mathcal{E}$) and inhibitory ($\mathcal{I}$) wave components. The result can be written in matrix form, as we did in (6):

$$\mathcal{M}\mathbf{Y} = \mathbf{J}_0, \tag{13}$$

where $\mathbf{Y} = (\mathcal{E}, \mathcal{I})^T$ is the wave-component vector, $\mathbf{J}_0 = -j_0(\alpha, 1-\alpha)^T$ is the stimulation, and $\mathcal{M}$ is the matrix of wave-component interaction:

$$\begin{bmatrix} \mathcal{M}_{EE} & \mathcal{M}_{EI} \\ \mathcal{M}_{IE} & \mathcal{M}_{II} \end{bmatrix} = \begin{bmatrix} W_{EE} - \frac{3}{4}\gamma_E \mathcal{E}^2 - 1 - k^2 D_{EE} & -W_{EI} + k^2 D_{EI} \\ W_{IE} - k^2 D_{IE} & -W_{II} - \frac{3}{4}\gamma_I \mathcal{I}^2 - 1 + k^2 D_{II} \end{bmatrix}.$$

The latter matrix represents mutual influence of wave components. Two terms of the matrix — $\mathcal{M}_{EE}$ and $\mathcal{M}_{II}$ — depend on the amplitudes $\mathcal{E}$ and $\mathcal{I}$, indicating that the interaction of wave components depends on stimulus contrast (in addition to its dependence on $k$ and on weights of connections between cells). This dependence on the amplitudes of wave components is negligible at low contrasts (where terms that include $\gamma_E$ and $\gamma_I$ are very small) and it becomes progressively more prominent as $\mathcal{E}$ and $\mathcal{I}$ increase with contrast.

It is useful to distinguish between two kinds of coefficients describing circuit interactions. On the one hand, coefficients $W$ and $D$ in (1–2) depend solely on weights of synaptic connections between cells (henceforth "cell interaction coefficients") that describe stimulus-independent interaction of cells in the circuit. On the other hand, the terms $\mathcal{M}_{EE}$, $\mathcal{M}_{EI}$, $\mathcal{M}_{IE}$, $\mathcal{M}_{II}$ in (14) describe the interaction of components of neural waves rather than the interaction of cells. These coefficients $\mathcal{M}$ (henceforth "wave interaction coefficients") depend both on cell interaction coefficients and on wave properties (such as wave frequency and amplitude). As a result, wave interaction coefficients depend on stimulus frequency and contrast.

To investigate effects of stimulus contrast, we rewrite (13) as

$$\mathbf{Y}(\mathbf{J}_0) = \mathcal{M}^{-1}\Big(\mathbf{Y}(\mathbf{J}_0)\Big)\mathbf{J}_0, \tag{14}$$

where $\mathcal{M}^{-1}(\mathbf{Y})$ is an inverse matrix. Equation (14) can be solved iteratively:

$$\mathbf{Y}_{n+1} = \mathcal{M}^{-1}(\mathbf{Y}_n)\mathbf{J}_0,$$

where $n$ is the iteration step number. The iterative procedure can be written explicitly:

$$\mathcal{I}_{n+1} = j_0 \frac{\Psi_E + \Phi_E k^2 + \eta_I \mathcal{I}_n^2}{(k^2 - k_n^2 + \lambda^2 + \xi_I \mathcal{I}_n^2 - \xi_E \mathcal{E}_n^2)^2 + 4\lambda^2 k_n^2 + \sigma_I \mathcal{I}_n^2 - \sigma_E \mathcal{E}_n^2},$$

$$\mathcal{E}_{n+1} = j_0 \frac{\Psi_I + \Phi_I k^2 + \eta_E \mathcal{E}_n^2}{\underbrace{\left(k^2 - k_n^2 + \lambda^2 + \xi_I \mathcal{I}_n^2 - \xi_E \mathcal{E}_n^2\right)^2}_{\text{Term A}} + \underbrace{4\lambda^2 k_n^2 + \sigma_I \mathcal{I}_n^2 - \sigma_E \mathcal{E}_n^2}_{\text{Term B}}}, \tag{15}$$

where constants $\Psi_E$, $\Psi_I$, $\Phi_E$, and $\Phi_I$ are functions of cell interaction coefficients. Constants $\eta_E$, $\eta_I$, $\xi_E$, $\xi_I$, $\sigma_E$ and $\sigma_I$ are also functions of cell interaction coefficients; they are proportional to the nonlinearity parameters $\gamma_E$ and $\gamma_I$. A zero of Term A in the denominator of (15) corresponds to the condition of system resonance at zero temporal frequency (similar to the minimum of $H$ in Equation 8). And Term B in the denominator of (15) determines the extent of selectivity, which is the "width" of tuning to spatial frequency. Equation 15 allows one to derive results of the $(n+1)$th iterations for $\mathcal{E}_{n+1}$ and $\mathcal{I}_{n+1}$ using results of the $n$th iterations for $\mathcal{E}_n$ and $\mathcal{I}_n$. We used the iterative procedure defined by (15) to produce Figure 8A-C.

Conditions of spatial resonance in the circuit, where Term A in the denominator of (15) is zero, are:

$$k_r^2 = (k_n^2 - \lambda^2) - \xi_I \, \mathcal{I}^2 + \xi_E \, \mathcal{E}^2, \tag{16}$$

The term $k_n^2 - \lambda^2$ in (16) determines the resonance frequency $k_r$ at low contrasts. Notice that $k_r$ and $k_n$ could coincide when wave damping in the system is low (i.e., when $\lambda$ is small). Notice also that the terms for $\mathcal{E}$ and $\mathcal{I}$ in (16) have opposite signs. To appreciate this result, suppose that excitatory and inhibitory waves have similar magnitudes, $\mathcal{E} = \mathcal{I} = \mathcal{A}$. We can therefore rewrite (16) as

$$k_r^2 = (k_n^2 - \lambda^2) - \mathcal{A}^2(\xi_I - \xi_E). \tag{17}$$

Given that the terms $k_n^2 - \lambda^2$, $\xi_I$, and $\xi_E$ are constants, and that $\mathcal{A}$ is an increasing function of contrast, resonance frequency $k_r$ will decrease with contrast when $\xi_I > \xi_E$ (Figure 8C) and increase with contrast when $\xi_I < \xi_E$ (Figure 8B).

# References

Dobkins, K. R. and Albright, T. D. (1994). What happens if it changes color when it moves?: the nature of chromatic input to macaque visual area MT. *Journal of Neuroscience*, 14(8):4854–4870.

Li, X., Lu, Z.-L., Xu, P., Jin, J., and Zhou, Y. (2003). Generating high gray-level resolution monochrome displays with conventional computer graphics cards and color monitors. *Journal of Neuroscience Methods*, 130(1):9–18.

Pawar, A. S., Gepshtein, S., Savel'ev, S., and Albright, T. D. (2019). Mechanisms of spatiotemporal selectivity in cortical area MT. *Neuron*, 101(3):514–527.

Savel'ev, S. and Gepshtein, S. (2014). Neural wave interference in inhibition-stabilized networks. In Gencaga, D., editor, *Proceedings of the First International Electronic Conference on Entropy and its Applications*, page c002.

Wilson, H. R. and Cowan, J. D. (1972). Excitatory and inhibitory interactions in localized populations of model neurons. *Biophysical Journal*, 12(1):1–24.

# Supplementary Materials

Supplemental Figure 1. Resampling analysis of spatial frequencies of lateral activation in human subjects

Supplemental Figure 2. Numbers of MT neurons measured to construct Figure 6

Supplemental Figure 3. Numbers of MT neurons measured to construct Figure 7

Supplemental Modeling Methods. Continuous model of a spatially distributed Wilson-Cowan circuit

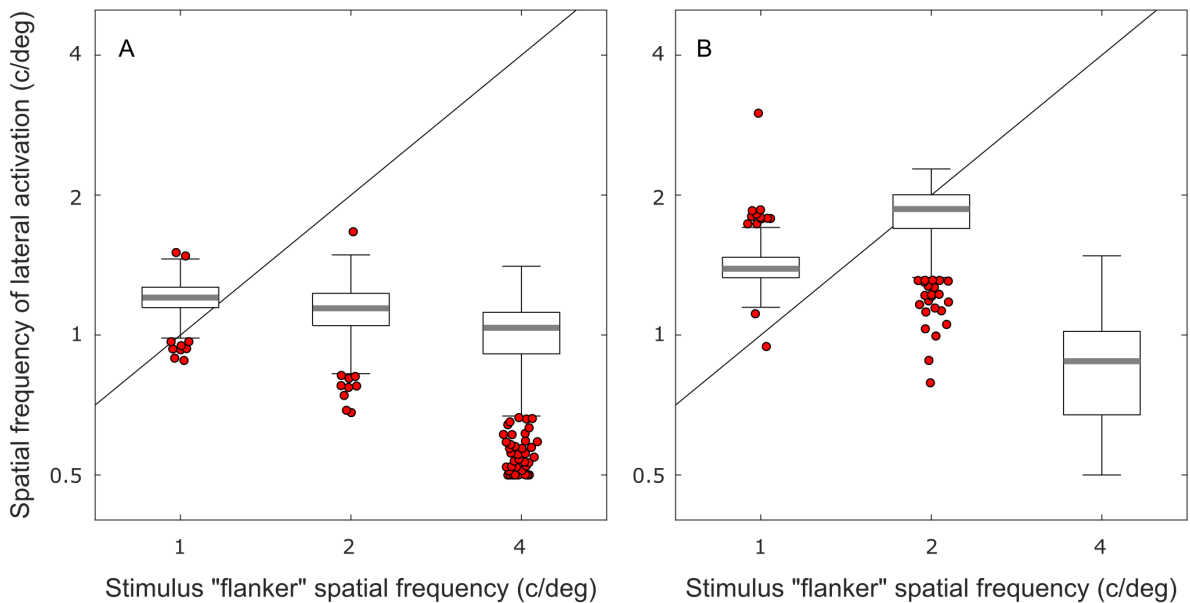

**Supplemental Figure 1. Resampling analysis of spatial frequencies of lateral activation in human subjects**

The boxplots represent results of resampling analysis of spatial frequencies (SF) of lateral activation waveforms fitted to contrast sensitivity data for two human subjects (Figure 3C). Results are shown for Subject 1 in panel A and for Subject 2 in panel B. Three values of stimulus SF in the abscissa of each panel correspond to three flanker SFs displayed on top of each column of plots in Figure 3C. The black line in each plot represents the SF of lateral activation that would have been observed if lateral activation had the same SF as stimulus SF.

Estimates of contrast sensitivity for each flanker SF were resampled with replacement 500 times. On each iteration of resampling, a harmonic function tapered by an exponential envelope was fitted to contrast sensitivity estimates for six probe-to-flanker distances (displayed in Figure 3C). Each boxplot represents a distribution of estimates of the SF of lateral activation: the horizontal gray bar marks the median, the top and bottom edges of the box mark the interquartile range (IQR), the whiskers mark the range of 1.5×IQR, and red dots represent the outliers.

For Subject 1, differences between estimates of lateral SF obtained at different values of stimulus SF were statistically insignificant ($p$=0.34 for stimulus SFs of 1 and 2 c/deg; $p$=0.26 for stimulus SFs of 2 and 4 c/deg). For Subject 2 the difference between estimates of lateral SF was statistically insignificant for stimulus SFs of 1 and 2 c/deg ($p$=0.07), yet lateral SF decreased significantly from stimulus SFs of 2 to stimulus SFs of 4 c/deg ($p \ll 0.01$). Parameters of the waveforms displayed in Figure 3C were obtained by averaging results of resampling.

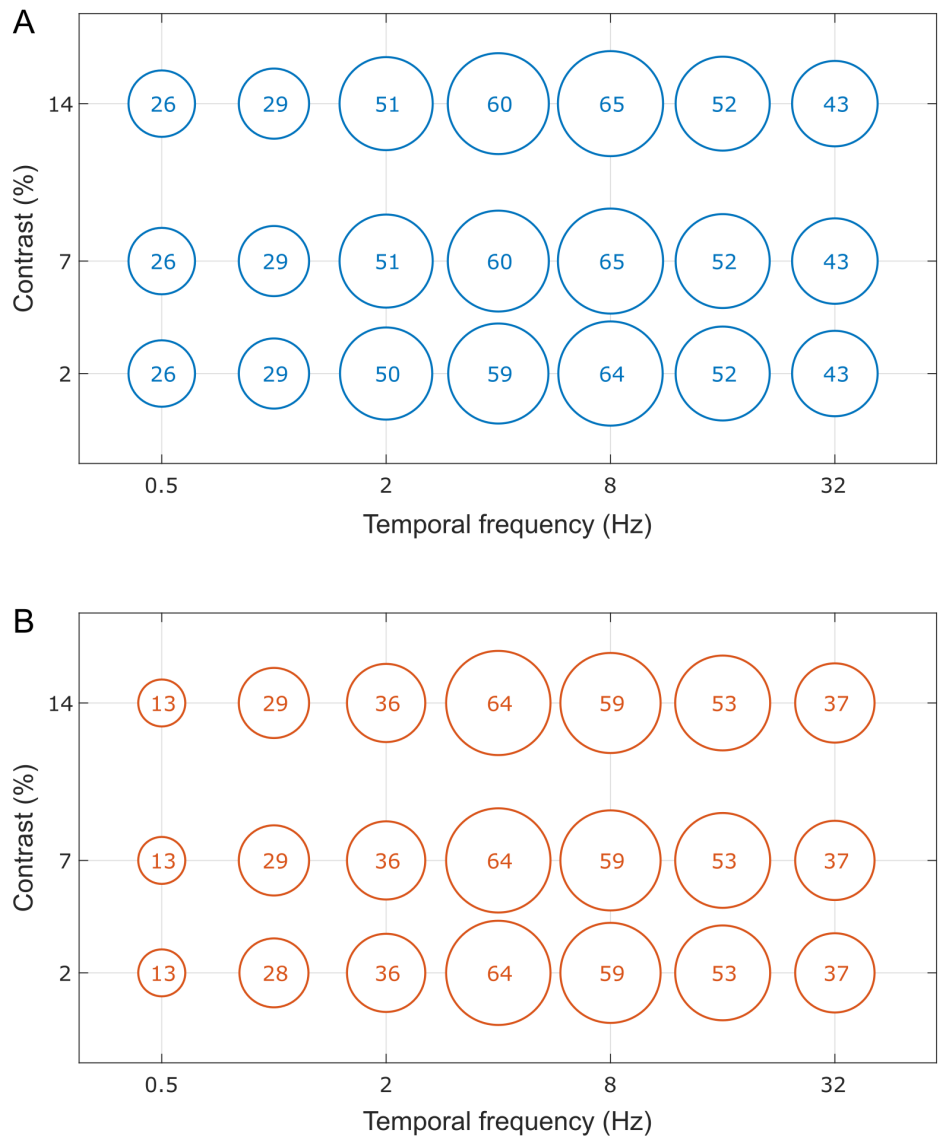

**Supplemental Figure 2. Numbers of MT neurons measured to construct Figure 6**

Circle sizes represent numbers of neurons measured in populations of MT neurons in Monkey 1 (A) and Monkey 2 (B) to obtain temporal response functions, such as those shown in the top row of plots in Figure 6. Each temporal response function was measured at seven stimulus temporal frequencies, separately for three stimulus contrasts (shown in this figure) and eight stimulus spatial frequencies (not shown here since there was little variation of neuron counts across stimulus spatial frequencies).

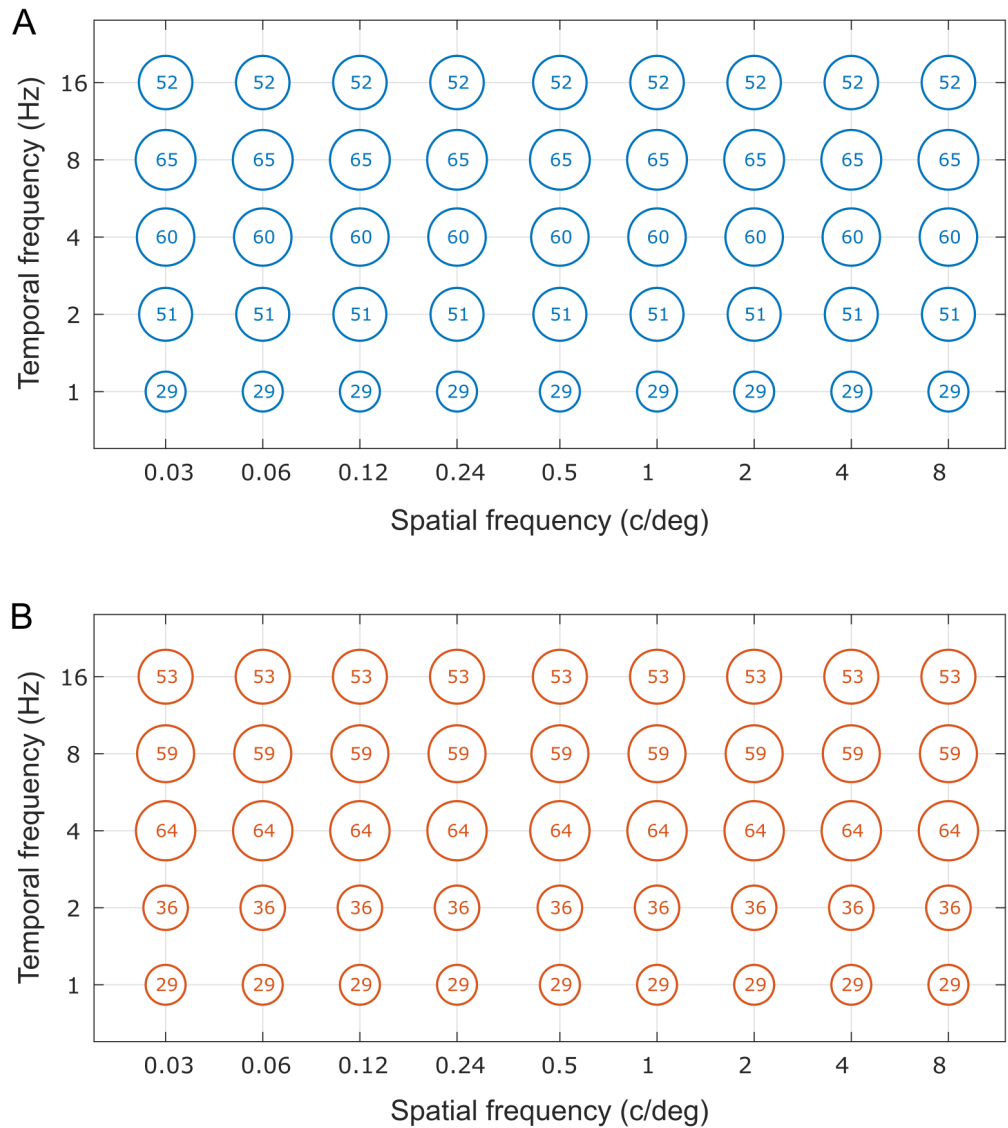

**Supplemental Figure 3. Numbers of MT neurons measured to construct Figure 7**

Circle sizes represent numbers of neurons measured in populations of MT neurons in Monkey 1 (A) and Monkey 2 (B) to obtain spatial response functions, such as those shown in the left column of plots in Figure 7. Each spatial response function was measured at nine stimulus spatial frequencies, separately for five stimulus temporal frequencies (shown in this figure) and three stimulus contrasts (not shown here since neuron counts for different stimulus contrasts were the same).

## Supplemental Modeling Methods

*Continuous model of spatially distributed Wilson-Cowan circuit*

Consider a system of Wilson-Cowan equations for a chain of identical network motifs, each containing an excitatory cell ("*E*") and an inhibitory cell ("*I*"), as shown in Figure 1A. This network architecture is described by the following system of equations:

$$\tau_E \frac{dr_E(l)}{dt} = -r_E + g_E(\mathcal{W}_E),$$
$$\frac{dr_I(l)}{dt} = -r_I + g_I(\mathcal{W}_I). \tag{1}$$

The variables $r_E$ and $r_I$ represent the firing rates of the excitatory and inhibitory cells, $\tau_E$ represents the relaxation time of the excitatory cell (in units of the relaxation time of the inhibitory cell), $l$ is a discrete index of location in the chain, and $g_E$, $g_I$ are sigmoid functions. The terms $\mathcal{W}_E$ and $\mathcal{W}_I$ represent sources of cell activation at each location:

$$\mathcal{W}_E = \Big[ w_{EE} r_E(l) + \bar{w}_{EE} r_E(l+1) + \bar{w}_{EE} r_E(l-1) \Big] - \Big[ w_{EI} r_I(l) + \bar{w}_{EI} r_I(l+1) + \bar{w}_{EI} r_I(l-1) \Big] + i_E(l,t),$$

$$\mathcal{W}_I = \Big[ w_{IE} r_E(l) + \bar{w}_{IE} r_E(l+1) + \bar{w}_{IE} r_E(l-1) \Big] - \Big[ w_{II} r_I(l) + \bar{w}_{II} r_I(l+1) + \bar{w}_{II} r_I(l-1) \Big] + i_I(l,t),$$

where $w$ and $\bar{w}$ represent the weights of connections respectively within and between the motifs. We assume a fully interconnected nearest-neighbor network. That is, each inhibitory (excitatory) cell is connected to the excitatory (inhibitory) cell within motifs, and it is also connected to inhibitory and excitatory cells in the nearest motifs. The inter-motif connections allow activation to propagate through the chain.

We assume that cell connections do not vary across location. That is, we disregard effects of spatial disorder and asymmetry. We also assume that stimulus inputs $i_E, i_I$ and cell responses $r_E, r_I$ vary slowly on the scale of inter-motif distance, so that the spatial period of the stimulus is much larger than the inter-motif distance. For this case, we can replace the discrete index of chain location $l$ by a continuous spatial variable $x$, to obtain:

$$r_E(l \pm 1) = r_E(l) \pm \frac{\partial r_E}{\partial x} + \frac{1}{2} \frac{\partial^2 r_E}{\partial x^2},$$

$$r_I(l \pm 1) = r_I(l) \pm \frac{\partial r_I}{\partial x} + \frac{1}{2} \frac{\partial^2 r_I}{\partial x^2}.$$

We consider the conditions of stimulation in which the response does not approach the saturation value of the sigmoid activation function, so that only weak deviations from the linear response are possible. In this case, by expanding the inverse sigmoid functions $g_E^{-1}$ and $g_I^{-1}$ in the Taylor series up to the third order, we obtain from the above the following pair of coupled partial differential equations:

$$W_{EE}r_E(x,t) + D_{EE}\frac{\partial^2 r_E}{\partial x^2} - W_{EI}r_I(x,t) - D_{EI}\frac{\partial^2 r_I}{\partial x^2} + \alpha j(x,t)$$

$$= g_E^{-1}\left(\frac{\partial r_E}{\partial t} + r_E\right) \approx \left(\frac{\partial r_E}{\partial t} + r_E\right) + \beta_E\left(\frac{\partial r_E}{\partial t} + r_E\right)^2 + \gamma_E\left(\frac{\partial r_E}{\partial t} + r_E\right)^3,$$

$$W_{IE}r_E(x,t) + D_{IE}\frac{\partial^2 r_E}{\partial x^2} - W_{II}r_I(x,t) - D_{II}\frac{\partial^2 r_I}{\partial x^2} + (1-\alpha)j(x,t)$$

$$= g_I^{-1}\left(\frac{\partial r_I}{\partial t} + r_I\right) \approx \left(\frac{\partial r_I}{\partial t} + r_I\right) + \beta_I\left(\frac{\partial r_I}{\partial t} + r_I\right)^2 + \gamma_I\left(\frac{\partial r_I}{\partial t} + r_I\right)^3, \quad (2)$$

where

- $W$ are the "interaction constants" reflecting the strengths of connections between the excitatory and inhibitory parts of the network ($W_{EE} = w_{EE} + 2\tilde{w}_{EE}$, $W_{EI} = w_{EI} + 2\tilde{w}_{EI}$, etc),

- and $D$ are the "diffusion constants" reflecting the strengths of excitatory and inhibitory connections between the motifs ($D_{EE} = \tilde{w}_{EE}$, $D_{EI} = \tilde{w}_{EI}$, etc) responsible for spatial propagation of excitatory and inhibitory influences through the network.

In the equation set (2), parameter $\alpha$ describes how the input current $j$ is divided between the excitatory and inhibitory cells. That is, in a system activated by the spatiotemporal stimulus $j(x,t)$, the input currents are $i_E(x,t) = \alpha j(x,t)$ and $i_I(x,t) = (1-\alpha)j(x,t)$.

The second- and third-order Taylor expansion coefficients, $\beta$ and $\gamma$, can be different for excitatory ($\beta_E, \gamma_E$) and inhibitory ($\beta_I, \gamma_I$) activation functions $g_E$ and $g_I$; they define the degree of system nonlinearity. The terms that include $\gamma$ control the part of response at the same frequency as the stimulus, while the terms that include $\beta$ control the part responsible for higher harmonics. Since in this work we are interested in the former part of response, we keep $\gamma$ terms and ignore $\beta$ terms, in the interest of making results of our analysis more tractable.

Equations (2) are key for our analysis of circuit response to spatially distributed, temporally extended stimuli.

*Model of distributed computation in two spatial dimensions*

In the previous section we investigated a model of one-dimensional (1D) neural chain with nearest-neighbor coupling. Such a model can describe interactions of stimuli shaped as parallel stripes or lines. To its advantage, the 1D model can be studied analytically, at least in part, greatly simplifying the task of finding the parameters that describe qualitatively different regimes of network function.

A more general model should be able to predict responses to two-dimensional (2D) stimuli. A model of higher-dimensional circuit will allow one to investigate different network geometries and manners of node coupling, offering a new and very interesting domain for analysis and simulation. Here we present first steps in the analysis of a 2D neural array, aiming to demonstrate how our approach of neural wave interference applies to such systems and how it may help interpretation of experimental results.

We consider a square lattice of nodes each containing one excitatory neuron and one inhibitory neuron connected as in Figure 1a in the main text. The nodes are connected along the *sides and diagonals* of each square cells of the lattice. The relative strengths of coupling along the sides and diagonals is controlled by parameter $\beta$, which is assumed to be the same for all the nodes. This model can be written as

$$
\begin{aligned}
\tau_E \frac{dr_E(l,m)}{dt} &= -r_E(l,m) \\
&+ g\left( w_{EE} r_E(l,m) + \bar{w}_{EE} \sum_E r_E - w_{EI} r_I(l,m) - \bar{w}_{EI} \sum_I r_I + i_E(l,m,t) \right), \\
\frac{dr_I(l,m)}{dt} &= -r_I(l,m) \\
&+ g\left( w_{IE} r_E(l,m) + \bar{w}_{IE} \sum_E r_E - w_{II} r_I(l,m) - \bar{w}_{II} \sum_I r_I + i_I(l,m,t) \right),
\end{aligned}
$$

(3)

where

$$
\begin{aligned}
\sum_E r_E = &\, r_E(l+1,m) + r_E(l-1,m) + r_E(l,m+1) + r_E(l,m-1) + \\
&+ \beta[r_E(l+1,m+1) + r_E(l+1,m-1) + r_E(l-1,m+1) + r_E(l-1,m-1)]
\end{aligned}
$$

and

$$
\begin{aligned}
\sum_I r_I = &\, r_I(l+1,m) + r_I(l-1,m) + r_I(l,m+1) + r_I(l,m-1) + \\
&+ \beta[r_I(l+1,m+1) + r_I(l+1,m-1) + r_I(l-1,m+1) + r_I(l-1,m-1)].
\end{aligned}
$$

If the firing rates vary slowly from one node to the next, we can employ a continuous approximation:

$$
\begin{aligned}
\tau_E \frac{\partial r_E}{\partial t} &= -r_E + W_{EE} r_E + D_{EE} \nabla^2 r_E - W_{EI} r_I - D_{EI} \nabla^2 r_I + i_E(x,y,t) \\
\frac{\partial r_I}{\partial t} &= -r_I + W_{IE} r_E + D_{IE} \nabla^2 r_E - W_{II} r_I - D_{II} \nabla^2 r_I + i_I(x,y,t)),
\end{aligned}
$$

(4)

where we use the standard notation of $\nabla^2 = \partial^2/\partial x^2 + \partial^2/\partial y^2$.

*Neural wave interference within an elliptical ring*

Just as in a 1D neural chain, the interference of neural waves generated in a 2D neural array may form a peculiar spatial pattern of excitation and inhibition. The pattern will consist of regions where the potential stimuli may be facilitated or suppressed by a high-contrast inducing stimulus applied elsewhere in the 2D array.

The facilitation and suppression can be revealed by measuring the contrast threshold of a faint (low contrast) probing stimulus placed at every location in the region of interest. Indeed, if such a probe appears in the facilitated or suppressed regions, its contrast threshold should be, respectively, lower or higher than in the absence of the inducer (as in Figure 3C in the main text). This is the approach pursued by Kovács & Julesz (1994) in a seminal psychophysical study of how contrast threshold of human observers was modulated inside contours of 2D visual shapes (e.g., Figure 9C in the main text).

Following the example of Kovács & Julesz, we studied the pattern of neural wave interference generated in our 2D model by a high-contrast stimulus shaped as an elliptical ring. Namely, we simulated the network response to the input $j(l, m) = 0$ for any $l, m$ except of $l, m$ satisfying the inequality

$$|\sqrt{l^2/R_1^2 + m^2/R_2^2} - 1| < \Delta R/\sqrt{R_1 R_2},$$

where $R_1$ and $R_2$ are semi-axes of the elliptical ring. The inputs $j(l, m)$ are represented by a dashed black ellipse in Figure 9B in the main text for $(R_1, R_2) = (21,14)$. For the purpose of illustration, the semi-axes $R_1$ and $R_2$ were multiples of the half-period $\mathcal{L} = 7$ of neural-wave oscillation (which we measured using a point stimulus, Figure 9A), in order to obtain salient effects of constructive and destructive interference.

The pattern of neural wave interference generated by this stimulus was obtained by numerical simulation of Equations 3. The result is shown in Figure 9B in the main text. We find that, for the stimulus semi-axes $R_1 = 3\mathcal{L}$ and $R_2 = 2\mathcal{L}$, the competition between suppression and facilitation produces regions of excitation in the two foci of the elliptical ring, in agreement with a striking psychophysical result of Kovács & Julesz (1994).

*Temporal dynamics*

We studied response of a single node in our model (Figure 1A in the main text) to time dependent input $j(t)$ split (as before) to excitatory and inhibitory parts: $i_E = \alpha j(t)$ and $i_I = (1 - \alpha)j(t)$, where $\alpha$ is a fraction of the input injected in the excitatory cell.

This results in the following set of equations (*cf.* Ozeki et al., 2009):

$$\tau_E \frac{dr_E}{dt} + r_E - w_{EE}r_E + w_{EI}r_I = \alpha j(t)$$

$$\frac{dr_I}{dt} + r_I - w_{IE}r_E + w_{II}r_I = (1 - \alpha)j(t). \tag{5}$$

These equations are identical to Equation 2 of main methods if we set all inter-node weights $\mathcal{W}$ to zero. The general solution of Equations 5 can be written by using Green's functions $G_E(t, t_0), G_I(t, t_0)$ in the following form:

$$r_E = \int_0^\infty G_E(t, t_0) j(t_0) dt_0, \quad r_I = \int_0^\infty G_I(t, t_0) j(t_0) dt_0, \qquad (6)$$

where the Green functions satisfy the following pulse-input equations:

$$\tau_E \frac{dG_E(t, t_0)}{dt} + (1 - w_{EE}) G_E(t, t_0) + w_{EI} G_I(t, t_0) = \alpha \delta(t - t_0)$$

$$\frac{dG_I(t, t_0)}{dt} + (1 + w_{II}) G_I(t, t_0) - w_{IE} G_E = (1 - \alpha) \delta(t - t_0). \qquad (7)$$

with $G_E(t < t_0, t_0) = G_I(t < t_0, t_0) = 0$. The Green function can have the shape of a damped oscillation:

$$G_E = C e^{-\gamma(t - t_0)} \cos(\omega_f(t - t_0) - \phi), \qquad (8)$$

where

$$2\gamma = (1 - w_{EE})/\tau_E + w_{II} + 1 > 0$$

and

$$4\tau_E^2 \omega_f^2 = 4\tau_E w_{EI} w_{IE} - (\tau_E(w_{II} + 1) - 1 + w_{EE})^2 > 0.$$

Parameters $\gamma$ and $\omega$ depend on weights of connections between cells, and parameters $C$ and $\phi$ depend on the weight of connections and on $\alpha$. Since the Green function in Equation 8 is similar to the weighting function obtained in psychophysical studies of human vision (Manahilov 1995; 1998), we used the function in Equation 8 in our analysis of duration threshold.

Using stimuli at different contrast and durations, we determined the profile of circuit response $r_E(t)$ by superposition (Equation 6) of the oscillations of $r_E$ elicited by successive parts of the stimulus. To estimate duration threshold for a fixed input $j_0$ we compared the maximum of the excitatory cell response $r_E$ with the threshold $r_c$, assuming that the stimulus is detected when the response exceeds $r_c$. Thus we obtained the equation for duration threshold, plotted in Figure 10D in the main text.

## References

[1] Kovács, I. and Julesz, B. (1994). Perceptual sensitivity maps within globally defined visual shapes. *Nature*, 370(6491): 644–646.

[2] Manahilov V (1995). Spatiotemporal visual response to suprathreshold stimuli. *Vision Research*, 35: 227–237.

[3] Manahilov V (1998). Triphasic temporal impulse responses and Mach bands in time. *Vision Research*, 38: 447–458.

[4] Ozeki H, Finn IM, Schaffer ES, Miller KD, Ferster D (2009). Inhibitory stabilization of the cortical network underlies visual surround suppression. *Neuron*, 62(4): 578–592.

[5] Tadin D, Lappin JS, Gilroy LA, Blake R (2003). Perceptual consequences of center-surround antagonism in visual motion processing. *Nature* 424: 312–315.

[6] Wilson, H. R. and Cowan, J. D. (1972). Excitatory and inhibitory interactions in localized populations of model neurons. *Biophysical Journal*, 12(1): 1–24.

[7] Wilson, H. R. and Cowan, J. D. (1973). A mathematical theory of the functional dynamics of cortical and thalamic nervous tissue. *Kybernetik*, 13(2): 55–80.